\documentclass[useAMS,usenatbib]{mn2e}
\usepackage{graphicx,fleqn,natbib,amssymb}
\usepackage{xcolor}

\DeclareGraphicsExtensions{.pdf,.cps,.ps,.pdf.gz,.ps.gz,.pdf.Z}
\usepackage{dcolumn}
\usepackage{pdflscape}
\usepackage{dblfloatfix}

\newcommand\textlcsc[1]{\textsc{\MakeLowercase{#1}}}

\title[V404 Cygni]{{\it Swift} UVOT observations of the 2015 outburst of V404 Cygni}

\author[Oates et al.]{S. R. Oates$^{1,2}$, S. Motta$^{3}$, A. P. Beardmore$^{4}$, D. M. Russell$^{5}$, P. Gandhi$^{6}$, 
\newauthor N. P. M. Kuin$^{7}$, M. De Pasquale$^{7,8}$, D. Altamirano$^{6}$,  A. A. Breeveld$^{7}$,  
\newauthor A. J. Castro-Tirado$^{2,9}$, C. Knigge$^{6}$, M. J. Page$^{7}$, D. Steeghs$^{1}$ \\
  $^{1}$ Department of Physics, University of Warwick, Coventry, CV4 7AL, UK; s.oates@warwick.ac.uk\\
  $^{2}$ Instituto de Astrofísica de Andaluc\'{i}a (IAA-CSIC), Glorieta de la Astronom\'{i}a s/n, E-18008, Granada, Spain\\
  $^{3}$ University of Oxford, Department of Physics, Astrophysics, Denys Wilkinson Building, Keble Road, OX1 3RH, Oxford, United Kingdom\\
  $^{4}$ X-ray and Observational Astronomy Group, Department of Physics \& Astronomy, University of Leicester, LE1 7RH, UK\\
  $^{5}$ New York University Abu Dhabi, P.O. Box 129188, Abu Dhabi, UAE\\ 
  $^{6}$ School of Physics and Astronomy, University of Southampton,
Southampton, SO17 1BJ, UK\\
  $^{7}$ Mullard Space Science Laboratory, University College London, Holmbury St. Mary, Dorking, Surrey, RH5 6NT, UK\\
  $^{8}$ Department of Astronomy and Space Sciences, Istanbul University, Beyaz{\i}t, 34119, Istanbul, Turkey\\
  $^{9}$ Unidad Asociada Departamento de Ingenier\'{i}a de Sistemas y Autom\'{a}tica, E.T.S. de Ingenieros Industriales, Universidad de M\'{a}laga, Spain\\
  }

\begin{document}

\date{Accepted...Received...}

\maketitle

\label{firstpage}

\begin{abstract} 
The black-hole binary, V404 Cygni, went into outburst in June 2015, after 26 years of X-ray quiescence. We observed the outburst with the Neil Gehrels {\it Swift} observatory. We present optical/UV observations taken with the {\it Swift} Ultra-violet Optical Telescope, and compare them with the X-ray observations obtained with the {\it Swift} X-ray Telescope. We find that dust extinction affecting the optical/UV, does not correlate with absorption due to neutral hydrogen that affects the X-ray emission. We suggest there is a small inhomogeneous high density absorber containing a negligible amount of dust, close to the black hole. Overall, temporal variations in the optical/UV appear to trace those in the X-rays. During some epochs we observe an optical time-lag of $(15 - 35)$\,s. For both the optical/UV and X-rays, the amplitude of the variations correlates with flux, but this correlation is less significant in the optical/UV. The variability in the light curves may be produced by a complex combination of processes. Some of the X-ray variability may be due to the presence of a local, inhomogeneous and dust-free absorber, while variability visible in both the X-ray and optical/UV may instead be driven by the accretion flow: the X-rays are produced in the inner accretion disc, some of which are reprocessed to the optical/UV; and/or the X-ray and optical/UV emission is produced within the jet.
\end{abstract}


\section{Introduction}
\label{intro}
Low-mass X-ray binaries (LMXBs) consist of a compact object, neutron star or black hole, and a lower mass companion star. Matter is stripped off the donor star, usually via Roche-lobe overflow, and is accreted onto the compact object via an accretion disk. Black hole LMXBs are mostly transient systems, i.e., they spend most of their time in a quiescent state, where accretion activity is low making them faint in the X-ray band. During this state, the optical luminosity is also low, which often allows measurements to be performed of the donor star that can determine the orbital period and the mass of the compact accretor \citep[see][ and references therein]{lop19}. Black hole LMXBs sporadically enter an outburst phase that can last weeks to years, during which their luminosity increases by several orders of magnitude and dramatic changes in spectral and fast-time variability properties can be observed. 

In June 2015, the LMXB V404 Cygni, also known as GS 2023+338, went into outburst \citep{GCN:17929} after 26 years of quiescence. This prototypical black hole binary (BHB) was the first system with a conclusive mass function \citep{cas92}. It hosts a black hole with mass $M_{BH}\sim9\;{\rm M_{solar}}$ and a donor star with mass $\sim0.7\,{\rm M_{solar}}$ of spectral type K3 III \citep{kha10}. The orbit is $6.473\pm0.001$ days \citep{cas92} and the inclination is 67$^{+3}_{-1}$ degrees \citep{kha10}. V404 Cygni was first discovered as an X-ray source by the Ginga satellite in 1989, during an outburst \citep{IAU4782}. This detection prompted examination of photographic plates that showed V404 Cygni had previously been in outburst in 1938 and 1956 \citep[][see also references therein]{ric89,wag91}. 

Compared to previous outbursts, the June 2015 outburst is unique in terms of the exceptional coverage both in time and wavelength by a multitude of space and ground-based facilities. The outburst was first detected by the Burst Alert Telescope \citep[BAT; 15-350\,keV][]{bar05} on-board the NASA Neil Gehrels {\it Swift} satellite, hereafter referred to as {\it Swift}, and shortly afterwards by the {\it FERMI}/GBM \citep{GCN:17932}. It was also detected at high energies by MAXI \citep{ATel:7646} and by {\it INTEGRAL} \citep{ATel:7693}. A large number of optical/IR observatories took photometric and spectroscopic measurements of the June 2015 outburst. \cite{kim16} alone report 85000 individual data points in the $B$, $V$, $R$ and $I$ bands. Radio and sub-mm observations were also reported \citep{tet17,cha17}. As such, V404 Cygni has been extensively studied over a broad range of wavelengths. 

Using observations taken in the X-ray band ($0.5-10$\,keV) with {\it Swift}, \cite{motta17} reported highly variable absorption local to the source. This was interpreted as the effect of a cold outflow ejected by the source itself, that could sustain an unstable, radiation dominated thick disk because of the high accretion rates reached. 
\textit{INTEGRAL}/IBIS-ISGRI observations of the system during the June 2015 outburst suggest that the X-ray spectrum of V404 Cygni is not intrinsically very different from the spectrum of a standard BHB when the effects of heavy absorption are accounted for (i.e., at energies above 25\,keV; \citealt{san17}).

In optical spectra, taken over the entire active phase, \cite{mun16} detected P-Cyg profiles, suggesting the presence of a sustained outer accretion disc wind in V404 Cygni. A disc wind had previously been observed in BHBs at X-ray wavelengths, e.g., GRO J1655−40 \citep{mil06}, but had never been observed before at optical wavelengths, although a disk wind has been recently noticed in spectra of V4641 Sgr \citep[see][and references therein]{mun18}. P-Cyg profiles were also observed by {\it Chandra} in X-ray spectra of V404 Cygni \citep{kin15}. \cite{kim16} reported the observations of optical oscillations on time-scales of 100\,s to 2.5 hours from V404 Cygni during the June 2015 outburst, while \cite{gan16} presented observations of rapid (sub-second) optical flux variability. \cite{gan16} state that their data initially showed steep power spectra dominated by slow variations on $\sim 100-1000$\,s, while near the peak of the outburst (June 26), persistent sub-second optical flaring also appeared, close in time to giant radio and X-ray flaring. The fast flares were stronger in the red, and could be explained as optically-thin synchrotron emission from a compact jet. Radio data collected at different frequencies revealed striking variability \citep[see e.g.,][]{tet17,fen19,Mil19}, characterised by hundreds of flares visible at different frequencies over a two-week period.

With its fast slewing capabilities, {\it Swift} was able to provide immediate follow-up using its two narrow field telescopes: X-ray Telescope \citep[XRT;][]{bur05} and Ultra-violet and Optical Telescope \citep[UVOT;][]{roming}. Detailed spectral analysis of the {\it Swift}/XRT observations during the brightest phase of the outburst of V404 Cygni have already been presented in \cite{motta17}. In this paper, we focus on the {\it Swift}/UVOT observations and their comparison with the X-ray behaviour observed simultaneously by the {\it Swift}/XRT. V404 Cygni also had a secondary outburst in December 2015. Since this later outburst has limited sampling we focus on the June 2015 outburst, but for completeness in the Appendix we give  a short description of the data from the second outburst and present the photometry. This paper is organized as follows. In \S~\ref{reduction} we discuss the data reduction and analysis. The main results are presented in \S~\ref{results}. Discussion and conclusions follow in \S~\ref{discussion} and \ref{conclusions}, respectively.  All uncertainties throughout this paper are quoted at 1$\sigma$, unless otherwise stated. Throughout this paper, we use a distance to V404 Cygni of $2.39\pm0.14$\,kpc as determined through parallax measurements by \cite{mil09}.

\section{Observations \& Data Reduction}
\label{reduction}
{\it Swift}/BAT first triggered on V404 Cygni at 18:31:38 UT on the 15th June 2015 \citep[MJD =57188.771976 UTC;][]{GCN:17929} and this was followed by a series of secondary triggers at 2.26, 2.65, 2.78, 4.79 and 10.76 days after the first trigger \citep[equivalent to $200$\,ks, $229$\,ks, $240$\,ks, $414$\,ks and $930$\,ks after the first trigger;][]{GCN:17944,GCN:17945,GCN:17946,GCN:17949,GCN:17963}. To prevent further triggers, V404 Cygni was entered into the on-board catalogue on the 22nd June 2015, which set a minimum trigger threshold for this source. However, the trigger threshold was exceeded at 10.76 days, resulting in another automatic slew to the source and an automatic increase in the BAT trigger threshold. Additional follow-up of V404 Cygni by {\it Swift} was therefore performed in response to a number of target of opportunity (ToO) requests (PIs: Altamirano, Beardmore, Cadolle Bel, Gandhi, Heinz, Kennea, Kuulkers, Middleton, Motta, Plotkin, Rodriguez, Sivakoff and Vasilopoulos). 

For the following analysis we correct all times to that at the barycentre. This results in a barycentric corrected, BAT trigger time $\rm T_0$ = 57188.773986 MJD (TDB), which we consider as the time the outburst began. 

As {\it Swift}\ slews to a number of targets in a given 95 min orbit to maximise its observing efficiency, a UVOT and XRT observation of a specific source on a given day is broken up into one or more `snapshots' of continuous exposure. Therefore the observed light curves are intermittent on long time scales, e.g several minutes to hours, but within this have periods of continuous observations e.g., $\sim 0.2 - 1.8\,{\rm ks}$ at few second resolution depending on the instrument, signal to noise and observing mode. Therefore, {\it Swift} observations can provide information on the overall behaviour of the outburst and some of the fast variability, but will have gaps in the observations during the main outburst when the source was highly variable.

\subsection{{\it Swift} UV/Optical Telescope}
\label{UVOTreduction}
{\it Swift}/UVOT began settled observations of V404 Cygni 184s after the initial BAT trigger \citep{GCN:17929}. {\it Swift}/UVOT monitored the outburst for the next 70 days ($\sim 6000$\,ks), until it returned to quiescence. V404 Cygni was initially detected in 5 of the 7 UVOT filters ({\it white}, {\it v}, {\it b}, {\it u}, {\it uvw1}) at $>2\sigma$, but as the source brightened it also became detectable in the {\it uvm2} and {\it uvw2} filters. 

{\it Swift}/UVOT also observed the field of V404 Cygni a number of times prior to 2015: observations were taken in the three optical and three UV filters on 26th April 2009; several observations were performed in the {\it uvw1} filter between the 3rd July 2012 and 16th September 2012; and on the 14th October 2013 and 2nd December 2013 observations were taken in the {\it u} filter. 

UVOT observations of V404 Cygni were taken in image and event modes\footnote{Image Mode: the data are recorded as an image accumulated over a fixed period of time; Event Mode: the arrival time and position are recorded for every photon detected.}. Before extracting count rates from the event lists, the astrometry was refined following the methodology in \cite{oates09}. In order to be compatible with the UVOT calibration \citep{poole, bre11}, we used an aperture of 5$\arcsec$ to obtain the source counts from both the event and image mode data. However, since it is recommended to use a smaller aperture when the count rate is low \citep{poole}, we used a 3$\arcsec$ radius below a threshold of 0.5 counts per second. In this case, the source count rates were corrected to a 5$\arcsec$ radius aperture using a table of aperture correction factors contained within the calibration. The background counts were obtained from a number of circular regions positioned in blank regions of the sky situated near to the source position. The count rates were obtained from the event and image lists using the {\it Swift} tools \textsc{uvotevtlc} and \textsc{uvotsource}, respectively. The UVOT data were processed using the {\it Swift}\ software, version 4.4 (from HEASOFT release 6.19), using the UVOT calibration files (20160321) released on 2016-Mar-21.

It has recently come to light that the UVOT detector is less sensitive in a few small patches\footnote{https://heasarc.gsfc.nasa.gov/docs/heasarc/caldb/swift/docs/
\\uvot/uvotcaldb\_sss\_01.pdf} for which a correction has not yet been determined. Therefore we checked to see if the source falls on any of these spots in any of our images and excluded 3 individual {\it uvw1} exposures for this reason. We also excluded {\it white} filter data between 5.8 and 9.2 days after the initial BAT trigger due to saturation.
      
A nearby star is reported to be 1.4" away from the position of V404 Cygni \citep{uda91}. This star may not be physically associated with the system, though \cite{mai17} recently proposed it may be an F2 star forming a triple system with V404 Cygni. The brightness of this star is $20.59\pm0.05$ and $18.90\pm0.02$ magnitudes in the B and V bands (Vega system), respectively, and was observed resolved with V404 Cygni in quiescence, when the flux of V404 Cygni was $20.63\pm0.05$ and $18.42\pm0.02$ magnitudes in the B and V bands (Vega system), respectively. In UVOT observations, this nearby star and V404 Cygni are not resolved, the emission from both combining to form a single point source. The photometry of V404 Cygni is therefore contaminated by flux from this nearby star. Since V404 Cygni is many times brighter in outburst than the quiescent phase, the nearby star will not strongly affect the photometry while in outburst. However, the contribution from this nearby star to the flux measurements will increase as V404 Cygni decreases in brightness. In quiescence, it is expected that this nearby star and V404 Cygni each will contribute about 50 per cent of the flux within the UVOT aperture (in the V and B filters). We therefore determine the nearby star's contribution to the photometry as V404 Cygni returns to quiescence, taking the lowest pre-2015 UVOT magnitudes of V404 Cygni as an upper limit to the flux from this contaminating star, namely an AB magnitude of $u=22.66\pm0.51$. We calculate that the nearby star contributes approximately $\leq4\%$, $\leq10\%$, $\leq25\%$ of the measured flux at ${\rm T_0}+$ 11.6, 18 and 23 days (${\rm T}_0+$ $1000$\,ks, $1600$\,ks and $2000$\,ks), to the $u$-band. In addition, the nearby star will be fainter in the UV bands than in the $u$-band since it has an F-type spectrum. 

The count rates from both the image and event mode data were converted to AB magnitudes using the UVOT zero points and to flux using the AB flux conversion factors \citep{bre11}. The UVOT observations are displayed in Fig. \ref{v404lc_multipanel}. To correct for interstellar extinction we use an $A_v=4.0\pm0.4$ \citep{cas93,hyn09} assuming the Milky Way extinction law given in \cite{pei92}. We note that more recent measurements of extinction along the line of sight has been performed by \cite{ito17}, with $3.0 < {\rm A_V} < 3.6$ and \cite{rah17} who measure ${\rm A_V} = 3.82\pm0.36$. These measurements are slightly lower than that measured by \cite{cas93} and \cite{hyn09}. If we had chosen to use a lower $\rm{A_V}$ value, it would have systematically decreased our measured magnitudes and luminosities, but it would not have any strong influence on our conclusions. Finally, the extinction corrected fluxes were converted to luminosity.

\subsection{{\it Swift} X-ray Telescope}
\label{XRTreduction}
Following the initial {\it Swift}/BAT trigger \citep{GCN:17929}, 
the XRT observed V404~Cygni for a total of 171.4\ ks over
70 days. 114.4\ ks of data were collected in Photon Counting (PC)
mode ($\rm{T_0}+\,<$23\,ks and $\rm{T_0}+\,>127$\,ks) and 57.0\ ks in Windowed Timing (WT) mode, with the latter used when the source was bright ($\rm{T_0}+\,$19\,ks until $\rm{T_0}+\,140$\,ks). In the analysis that follows the XRT data were processed using the {\it Swift}\ software, 
version 4.4 (from HEASOFT release 6.16), using the latest XRT
gain calibration files for this epoch (released on 2015-Jul-21).

The analysis of the central point source is made more challenging
by the presence of a sometimes significant dust scattered X-ray halo
\citep{vas16,hei16,bea16}. The effect of the halo is most notable
when a strong X-ray flare is followed by a drop in the source count
rate to below $\sim 100\ {\rm count\, s^{-1}}$ \citep[see appendix A of][]{motta17}.

\begin{landscape}
  \begin{figure}
    \centering
    \vspace{-2.5cm}\includegraphics[angle=-90, scale=0.8]{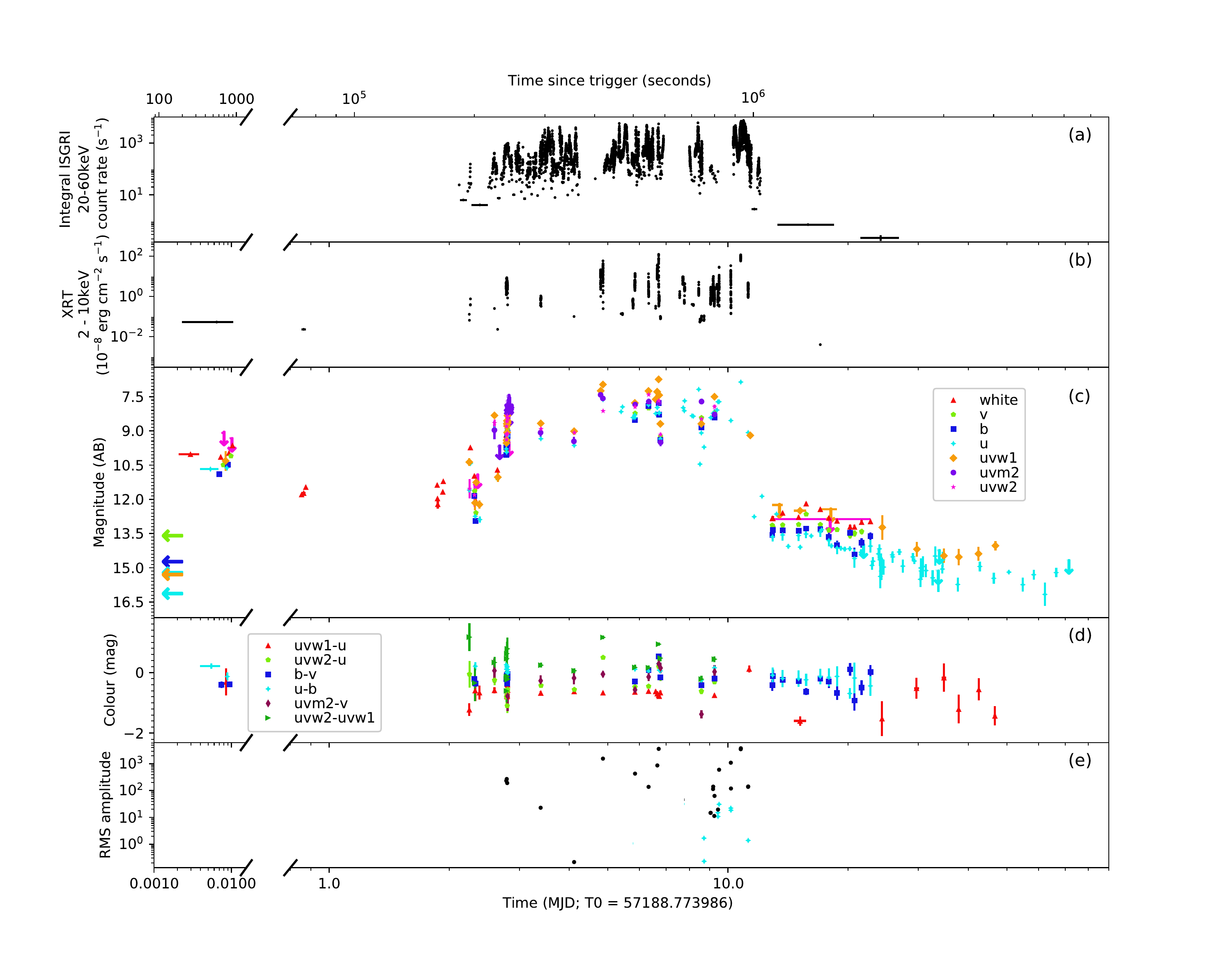}
    \vspace{-1.3cm}
  \caption{ Multi-wavelength observations of V404 Cygni performed during the June 2015 outbursts. $\rm T_0$ is taken as the time at which {\it Swift}/BAT first detected V404 Cygni on the 15th June 2015 and has been corrected to the barycenter. Panel a) shows the INTEGRAL $20-60$\,keV ISGRI light curve for the June 2015 outburst. Panel b) shows the X-ray $2-10$\,keV flux light curve for the June 2015 outburst. Panel c) displays the 7 optical/UV light curves observed by UVOT in image mode. In this panel, the left facing arrows represent pre-outburst observations performed by UVOT. Panel d) displays the colour light curves for several UVOT filters, determined only for pairs of points with S/N $>2$. Panel e) displays the excess variance determined from the power density spectra, for the X-ray (black) and $u$-band (light blue). Magnitudes are given in AB and have been corrected for the measured extinction $A_v=4.0$ \citep{cas93,hyn09}. Colours and symbols for each of the UVOT filters and colour light curves are provide in the legends. Downward arrows are $3\sigma$ upper limits computed when the signal to noise is $<2$. All times have been corrected to the barycentre.}
  \label{v404lc_multipanel}
  \end{figure}			
\end{landscape}
A $0.5-10.0$\,keV count rate light curve was created from the XRT
observations. The PC mode data were extracted at per
snapshot binning, using grade $0-4$ events. The PC data taken after day 14 were background subtracted using events contained within an annular region located inside the innermost dust scattering ring. WT
light curves were extracted at 1\,s binning from grade 0 events (as
this grade selection lessens the effects of pile-up, which becomes
important above $\sim 100-150\,{\rm count\,s^{-1}}$ in this mode).
When required, the halo contribution was estimated from the WT data
using events obtained within a $1-2$ arcminute-wide strip chosen to
lie as close to the central point source as possible when the halo
dominated. Appropriate corrections for (sometimes severe) pile-up
were also made to the data when necessary. This was achieved using
suitably sized annular extraction regions to excise the piled-up core
of the point spread profile, then correcting for the expected loss of
flux. The extraction regions thus ranged from circles of radius 23.6
arcseconds (when the central source was faint compared to the halo) to
annular regions of size $47.1-94.3$ arcseconds (when the central
source was extremely piled-up). Use of the $0.5-10.0$\,keV count 
rate light curve will be discussed in \S \ref{highres}.

To convert the X-ray count rate to luminosity we need to determine the count rate to flux conversion factors. This requires analysis of a large number of X-ray spectra, such as that performed by \cite{motta17}. \cite{motta17} computed X-ray fluxes in the energy range, $2-10$\,keV (note the narrower energy range compared to our $0.5-10.0$\,keV count rate light curve) using spectra extracted with 16\,s duration or long enough to contain about 1600 counts, resulting in 1054 data points with variable exposure. To compute these fluxes, the SEDs were fitted with an absorbed power law combined with a narrow Gaussian line around the 6.4\,keV Fe-K energy and interstellar absorption, when statistically required. A neutral absorber partially covering the source was also included when the fit showed improvement; see \cite{motta17} for further details. We use the $2-10$\,keV flux light curve from \cite{motta17} from this point onwards to produce the X-ray luminosity light curve, however, as \cite{motta17} only focused on the active part of the outburst (18th June - 26th June 2015), we also add data observed by {\it Swift}/XRT prior to and after this time, converting to flux following the same method. The XRT flux light curve, in the $2-10$\,keV energy range, is displayed in Fig. \ref{v404lc_multipanel}. We convert these 1080 flux points to luminosity. These luminosity measurements have been corrected for Galactic absorption, but they have not been corrected for intrinsic absorption as there may be some degeneracy between the absorption and reflection parameters in the $2-10$\,keV energy band \citep[see][]{motta17}, which if corrected for may give intrinsic luminosity values that may be significantly lower than the true values.

\subsection{High time resolution Analysis}
\label{highres}
\subsubsection{Light curves}
\begin{figure}
  \includegraphics[angle=0,scale=0.45]{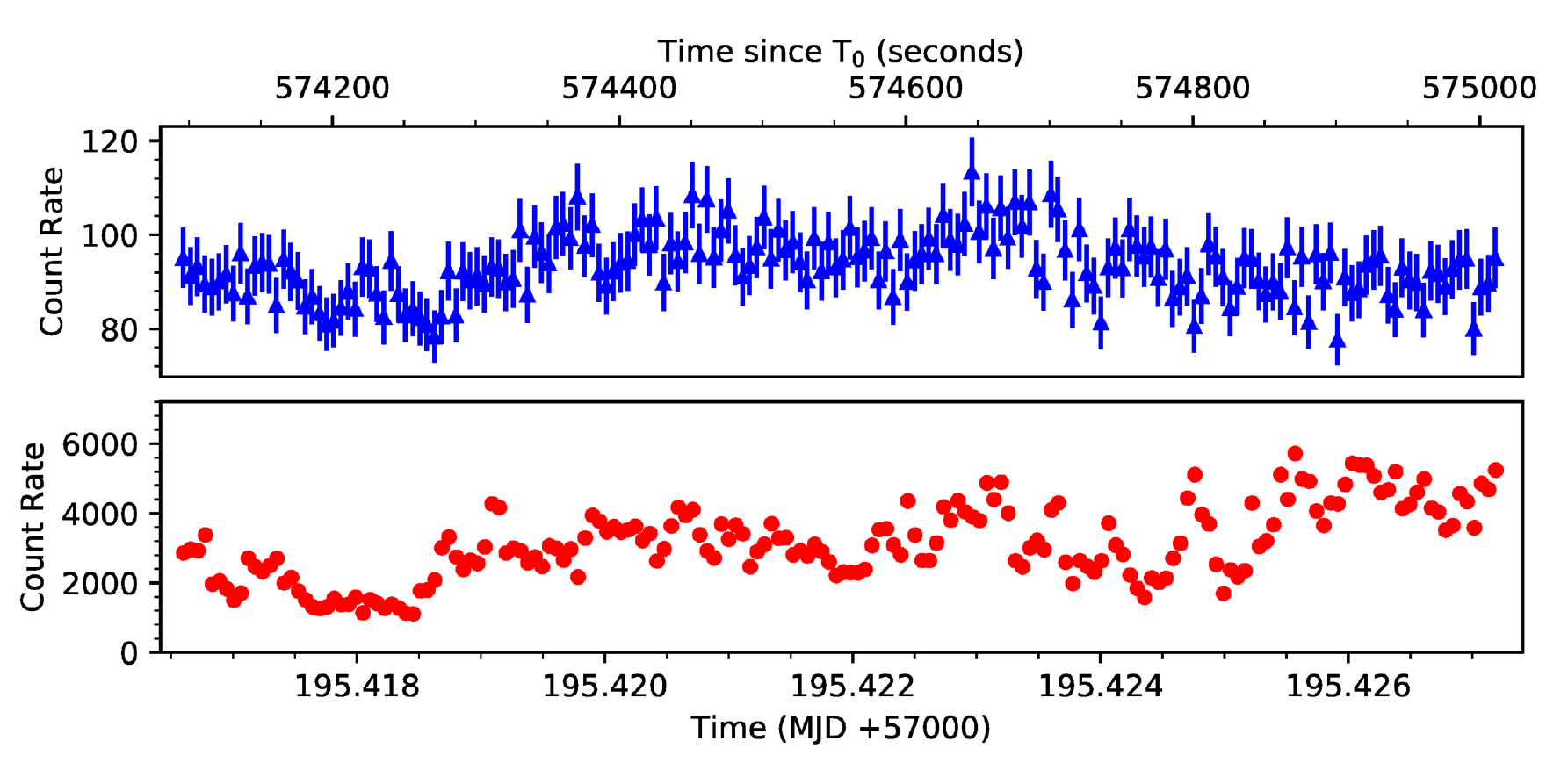}
  \includegraphics[angle=0,scale=0.45]{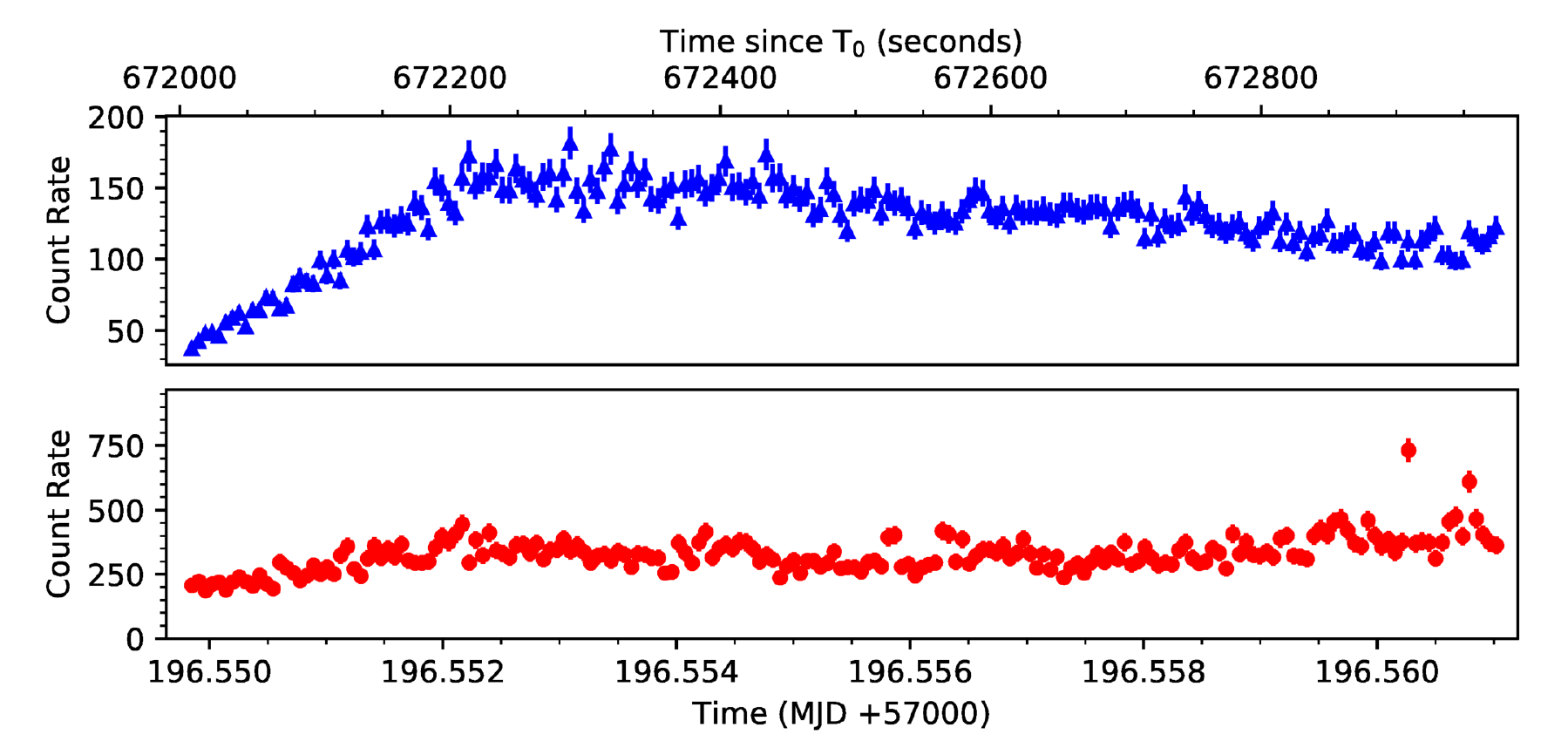}
  \includegraphics[angle=0,scale=0.45]{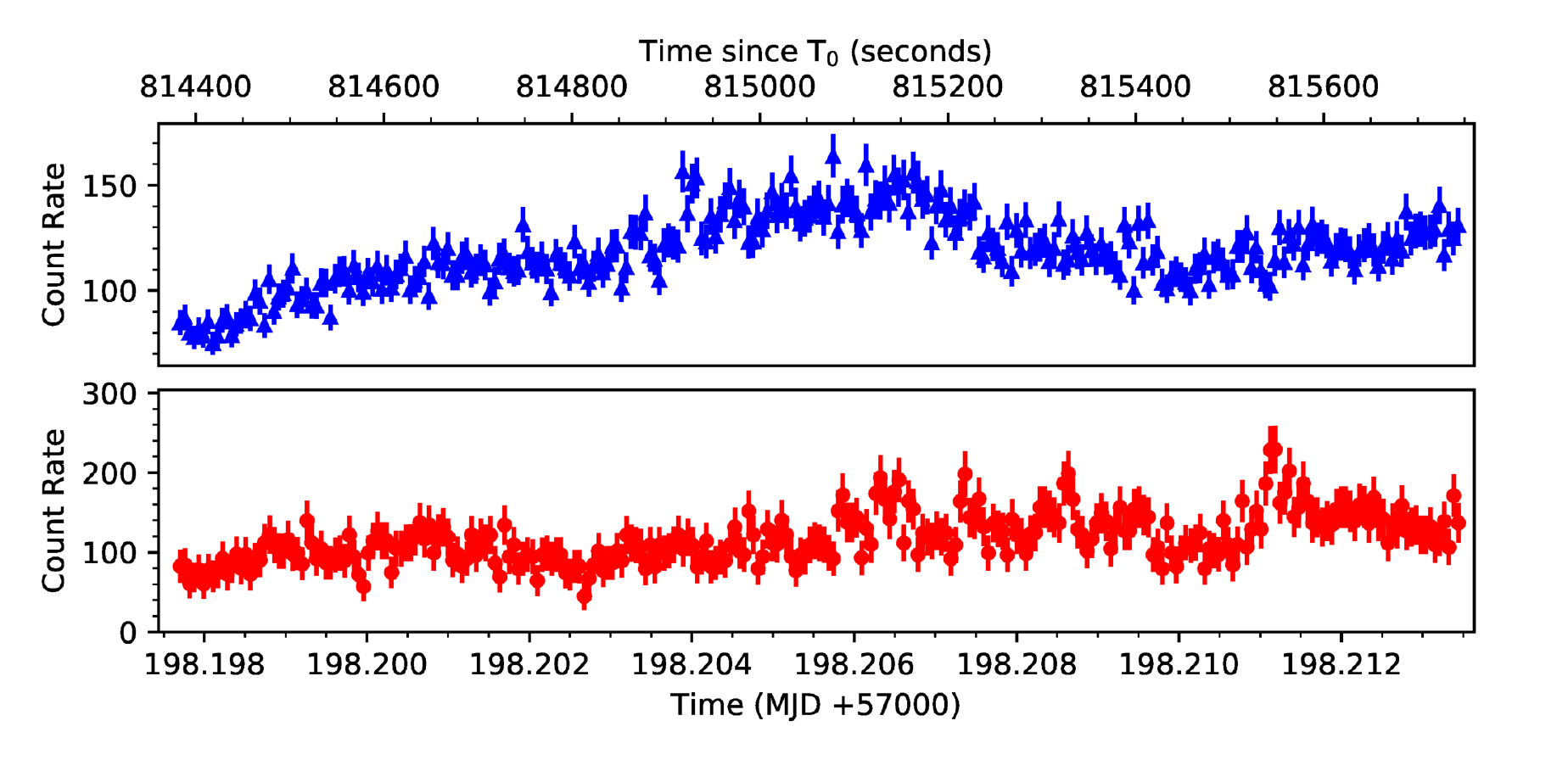}
  \caption{The UVOT $u$-band event mode observations (blue triangles) of V404 Cygni compared to the X-ray ($0.5-10$\,keV; red circles). In these panels, the X-ray data has been binned to have the same bin width as the $u$-band event mode data, which is binned uniformly with a bin width of 5\,s. These three panels were selected as examples of the type of temporal behaviour observed within V404 Cygni and how the behaviour compares in the two bands. The full sample is provided in the online material S.1.}
\label{eventlcXU}
\end{figure}

UVOT observations taken in event mode can be subdivided into smaller time bins in order to get a higher temporal resolution. Many of the UVOT $u$-band observations were observed in event mode. We extracted photometry from the UVOT $u$-band event mode data with a resolution of 5\,s. Since the time resolution of the X-ray light curves can also be resolved to a better time resolution than individual exposures and is observed at the same time, we can directly compare the behaviour in the X-rays with that in the $u$-band. We extract the $0.5-10$\,keV X-ray count rate in 1\,s intervals and then rebin the X-ray light curve so that it has the same time bins as the $u$-band data points. There are $\sim 4$k individual data points during the brightest phase of the outburst, i.e., ${\rm T_0}+\,240$\,ks $-$ 970\,ks (${\rm T_0}+\,2.8-11.2$ days). We display a selection of light curve segments from the resulting $u$-band and $0.5-10$\,keV X-ray count rate light curves in Fig. \ref{eventlcXU}. The full sample comparing the $u$-band with the X-ray $0.5-10$\,keV during the brightest phase of the outburst can be found in the on-line material, Fig. S.1, and also displayed in Fig. S.2. In Fig. S.1 we also provide the results from the X-ray spectral analysis by \cite{motta17} and as discussed in \S \ref{XRTreduction}, displaying the column density $N_H$, covering fraction and spectral index, $\Gamma$.

The X-ray count rate light curves have better temporal resolution compared with the X-ray luminosity light curves. In the following we therefore use the X-ray and $u$-band count rate light curves for the high time resolution temporal comparison with the $u$-band data. However, a caveat should be placed on the interpretation of the count rates, for instance, an increase in X-ray count rate does not necessarily mean an increase in X-ray flux at all X-ray energies. However, the X-ray count rate is a consistent photometric measure of the flux of X-rays through a fixed passband, and using this rather than luminosity, means that we do not introduce any systematics related to the different, longer time bin over which the spectra have been accumulated. 

\subsubsection{Timing}
\label{timing}
To determine if there is any time lag between the optical/UV and X-ray light curves, we calculate the cross-correlation for each continuous section of the 5\,s binned light curves. We determine the cross-correlation using the function {\it correlate} in the python numpy package. An example of the analysis performed on one section of the $u$-band and X-ray light curve is given in Fig. \ref{cross_corr}, while the results for the entire outburst can be found in the on-line material Fig. S.2. To determine if a given lag coefficient could be achieved due to chance, we perform a Monte Carlo simulation. For each section of light curve we simulate an X-ray light curve with the same power spectrum as the actual data. We perform a Fourier transform of the X-ray light curve and create a new X-ray light curve by randomising the Fourier phases, but maintaining the Fourier amplitude\footnote{We also tried varying the Fourier amplitude, using a Gaussian random sampling, taking the width of the Gaussian to be 10\% of the amplitude. We did not find any substantial changes to the 0.5, 2.5, 97.5 or 99.5 per cent values of the cross correlation distribution at a given lag time.}. With the simulated X-ray light curve we perform a cross correlation between those points and the corresponding, observed, $u$-band data. We repeated this $10^5$ times and store the resulting correlation coefficients for each lag time, for each iteration. Once all iterations are complete, for each lag time we extract values at $0.5, 2.5, 97.5$ and $99.5$ per cent from the resulting cross correlation distribution. 

To explore how the amplitude of the variability in the $u$-band and X-ray light curves changes with time, we measure the excess root mean square (rms) following the method outlined in \cite{vau03} and taking into account the measurement errors on the data. This method essentially takes a segment of uniformly binned data and determines the variance as the integral of the power spectral density (PSD) between two frequencies, subtracting the Poissonian noise from both the power spectra. We make use of the IDL routine \textlcsc{ts\_rmsflux.pro} and \textlcsc{DYNAMIC\_PDS.PRO} written by S. Vaughan\footnote{both IDL scripts are available at http://www.star.le.ac.uk/sav2/idl.html.}. We adapt the output so that the measurement errors are subtracted from the resulting variance and taken into account when determining the error on the variance. We use the full 1\,s X-ray ($0.5-10$\,keV) light curve (e.g., all observations, not only those overlapping with the $u$-band), rebinned to have a bin width of 5\,s, and the 5\,s $u$-band light curves. We calculate the PSD, and hence the excess rms, every 128 consecutive data points (i.e., 640\,s). The resulting light curve displaying the excess rms can be observed in the bottom panel of Fig. \ref{v404lc_multipanel}; we only display values where an excess could be determined above the error on the data, i.e., positive values of the excess. 

\section{Results}
\label{results}
The UVOT optical/UV image mode light curves for the June 2015 outburst of V404 Cygni are shown in Fig. \ref{v404lc_multipanel}. The temporal behaviour in the different optical/UV filters is similar. Over the course of observations V404 Cygni fluctuated in brightness by over 8 magnitudes. By the time UVOT began observing, V404 Cygni was already at a flux level $\sim 4$ magnitudes brighter than the level observed in quiescence during the pre-2015 observations. The optical/UV light curves rose to the highest level between $\rm T_0+\,3$ $-$ ${\rm T_0}+11$ days ($57191.5 - 57199.5$ MJD). At ${\rm T_0}+11$ days, the flux dropped rapidly, decreasing by 5 magnitudes within a day. Over the subsequent days the brightness continued to decrease, but at a slower rate, until it reached the level measured by {\it Swift}/UVOT prior to 2015. 

In Fig. \ref{v404lc_multipanel} we also display observations taken by {\it Swift}/XRT in the soft $2-10$\,keV X-ray band as well as those in the hard 25-60\,keV X-ray band performed by {\it INTEGRAL} IBIS/ISGRI \citep{jou17}. The IBIS/ISGRI light curve has been adaptively rebinned to achieve S/N $>8$ in each bin\footnote{http://www.isdc.unige.ch/integral/analysis\#QLAsources}  (see, e.g., \citealt{ATel:7758,ATel:8512,jou17,san17}, for details on the {\it INTEGRAL} data). The IBIS/ISGRI data has better sampling than the {\it Swift} optical/UV and X-ray observations, although from the three panels, it is apparent that the overall behaviour is the same in optical/UV through to hard X-rays. In particular, the drop back down to quiescence is observed in all of the top three panels of Fig. \ref{v404lc_multipanel} \citep[see, e.g.,][]{san17}. The soft/hard X-rays appear to vary strongly, flaring on time scales of minutes to hours. Due to the UVOT filter rotation, there are only a couple of images per optical/UV filter taken during these X-ray flares, but the optical/UV also seems to flare at similar times.

\subsection{Optical/UV Colour versus X-ray Column Density \& Photon Index}
To determine if there are changes in the optical/UV spectral shape during the outburst, we display the colour light curves for the UVOT filters in Fig \ref{v404lc_multipanel}. Since the UVOT observations in the different filters are not simultaneous, we take the difference in extinction-corrected AB magnitudes of the data points closest in time to each other. The typical difference in the start and stop time of each bin\footnote{where the start time is the start time of the earliest of the two exposures and the end time is the stop time of the latest exposure.} is 306\,s, but ranges from 44\,s to 1210\,s. At several points during observations the light curves show rapid changes in colour, for instance at $\rm{T}_0$+6.72 days ($\rm{T}_0$+581\,ks). These rapid changes are likely due to short timescale temporal evolution between the two filter visits since they are not strictly simultaneous. However, between $\rm{T}_0$+2 days and $\rm{T}_0$+10 days, the colour light curves are generally flat indicating no strong spectral changes throughout the main outburst. 

\begin{figure}
\centering
\hspace*{-0.4cm}
\includegraphics[angle=0, scale=0.55]{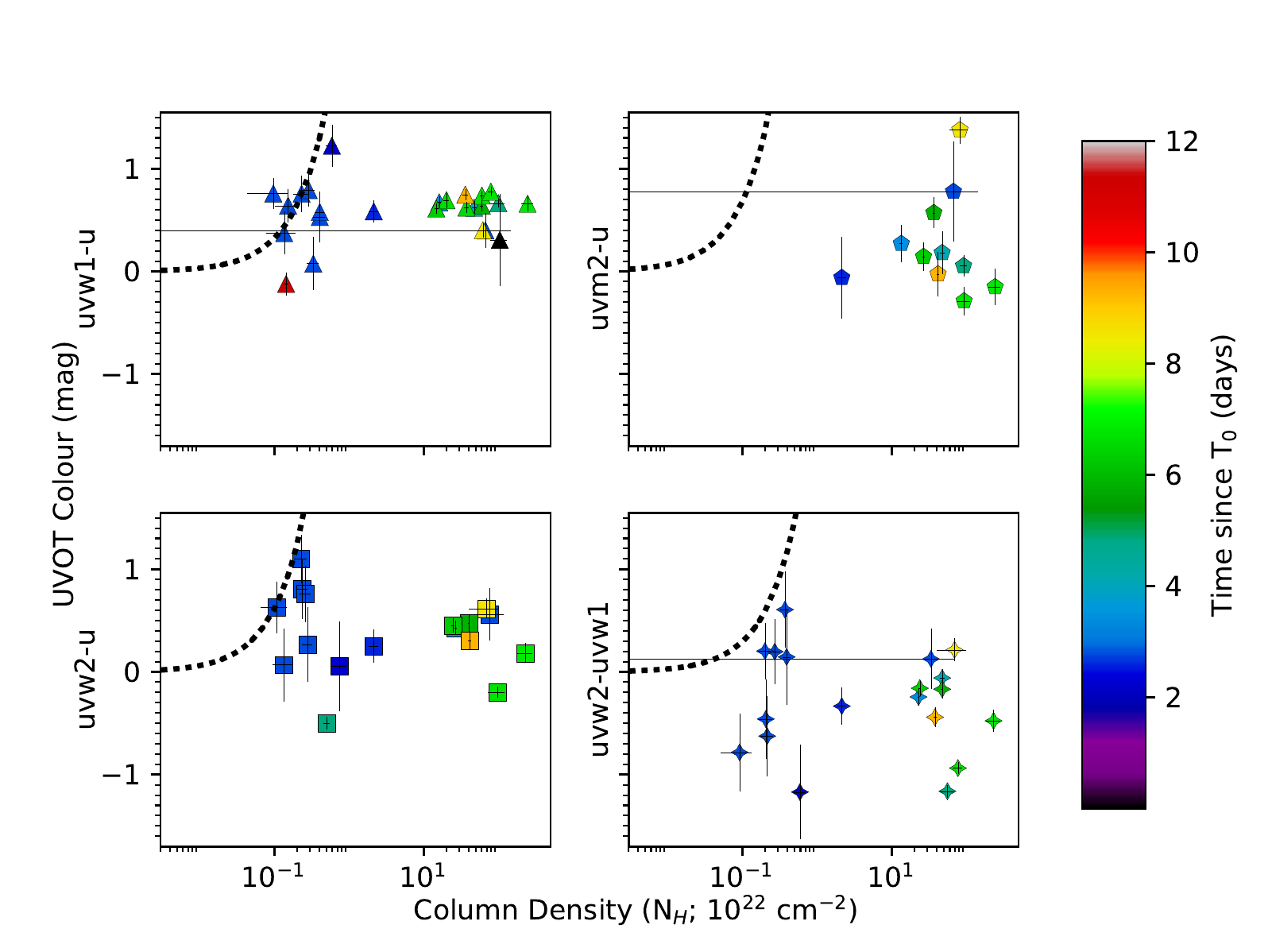}
\caption{UVOT colour versus column density of neutral hydrogen, 
$N_H$. We give 4 panels, comparing four different colour terms for the UVOT with $N_H$. The colour scale represents the time each point was determined. The black dotted line represents the expected relationship between $N_H$ and $E(\lambda_1-\lambda_2)$ for the Milky Way, where $\lambda_1$ and $\lambda_2$ represent the 2 different filters. Since extinction has a strong effect in the UV, the expected colour difference in the UV for a given $N_H$ will depend on the exact two filters. However while we observe $N_H$ to change significantly, and cluster into high and low $N_H$ groups, the UVOT colour does not become redder with increasing $N_H$. This suggests strong changes in absorption do not correspond to changes in optical/UV extinction.}
\label{colour-NH}
\end{figure}

In Fig \ref{colour-NH} we display UVOT colour versus the column density of neutral hydrogen, $N_H$, determined from the X-ray spectra as detailed in \S \ref{XRTreduction}. The figure consists of four panels, each displaying the difference in magnitude between a different pair of UV filters versus $N_H$. Since the colour points span a larger time interval than the interval typically used to extract the X-ray spectra, we take the average $N_H$ value of the spectra that fall within the start and stop time of the UVOT colour data point. In the determination of the average $N_H$ value, we do not include spectra for which the $N_H$ could not be constrained (247/1080), which may weight the average $N_H$ value to higher values. In Fig \ref{colour-NH}, we also show the relationship between extinction between the two filters, $E(\lambda_1-\lambda_2)$, and the column density of neutral hydrogen, $N_H$, as expected for the Milky Way. Since extinction has a stronger effect in the UV, the colour difference between an optical and a UV filter is expected to be larger than the $E(B-V)$ for a given value of $N_H$. While the $N_H$ value changes in time, clustering into high and low $N_H$ groups, the UVOT colour does not become redder with increasing $N_H$.

\begin{figure}
\centering
\hspace*{-0.4cm}
\includegraphics[angle=0, scale=0.45]{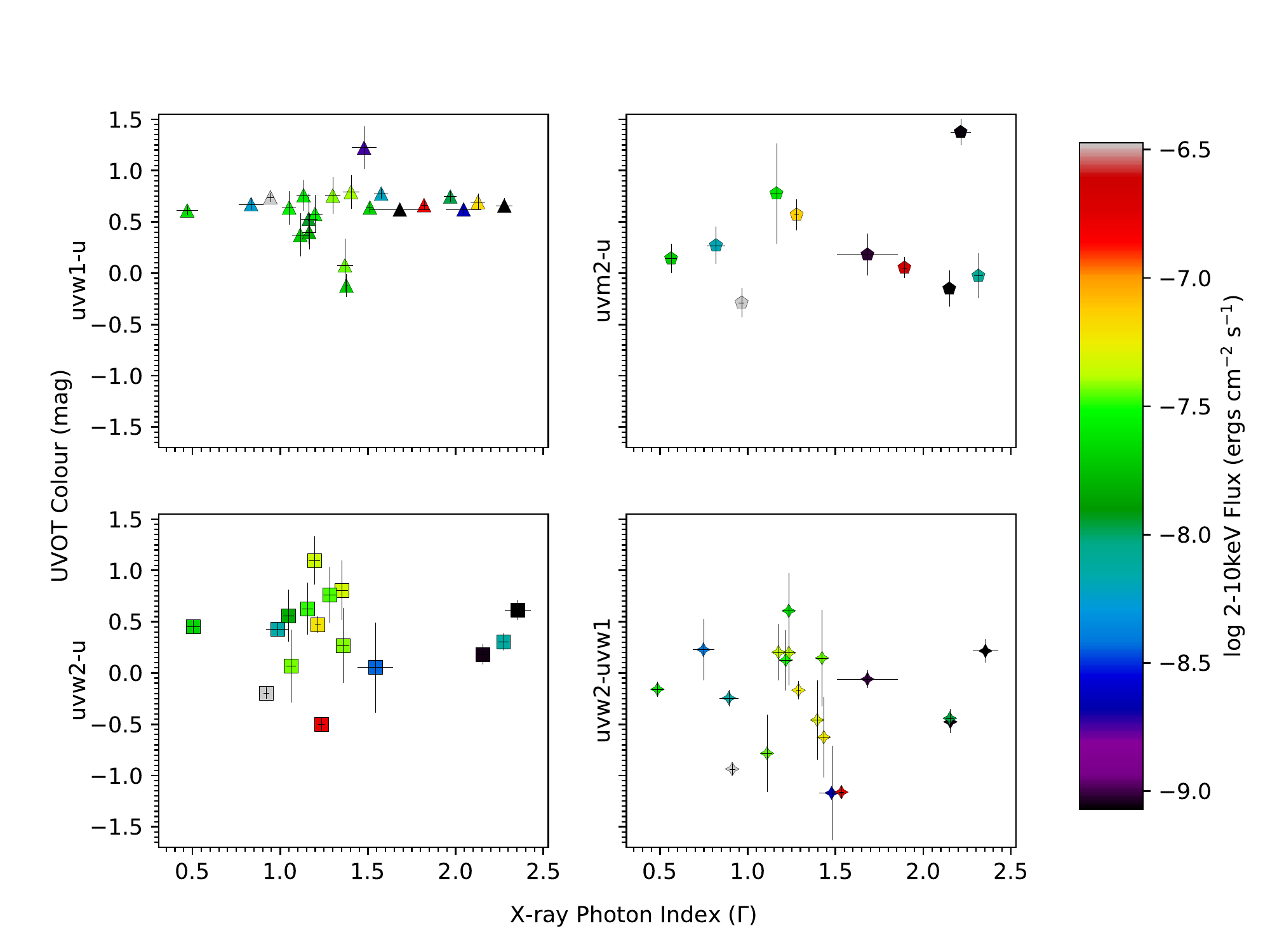}
\caption{UVOT colour versus the X-ray photon index, $\Gamma$. We give 4 panels, comparing four different colour terms for the UVOT with $\Gamma$. The colour scale represents the average X-ray flux during the time each colour point was determined. The bottom two panels show much wider range in colour, but the uncertainty is typically larger than the upper panels. However, for all 4 panels there is no apparent correlation between optical/UV colour and $\Gamma$, suggesting that changes in the X-ray spectrum do not correspond to a changes in the optical spectrum.}
\label{colour-gamma}
\end{figure}

In Fig. \ref{colour-gamma}, we display the same UVOT colours against X-ray photon index, $\Gamma$, in order to determine if a change in the X-ray spectrum corresponds to a change in the optical spectrum. Since the colour points span a larger time interval than the interval typically used to extract the X-ray spectra, we take the average $\Gamma$ value of the spectra that fall within the start and stop time of the UVOT colour data point. In the determination of the average $\Gamma$ value, we do not include spectra for which the $\Gamma$ could not be constrained (52/1080), which may weight the average $\Gamma$ value to softer values. \cite{motta17} note that spectra with $\Gamma <1$ are not consistent with the spectral slope observed by INTEGRAL, $\Gamma>1.4$ \citep{san17}. \cite{motta17} find the spectra become harder when a reflection component is included.

In the top two panels, displaying $uvw1-u$ and $uvm2-u$, the UVOT colours are relatively stable at $\sim 0.7$ while the X-ray $\Gamma$ varies from $\sim 0.5-2.5$ over the brightest phase of the outburst. The bottom two panels show a much wider range in colour, but the uncertainty is typically larger than the upper panels. In any case, in all 4 panels there is no apparent correlation between optical/UV colour and $\Gamma$.

\begin{figure}
\includegraphics[angle=0,scale=0.45]{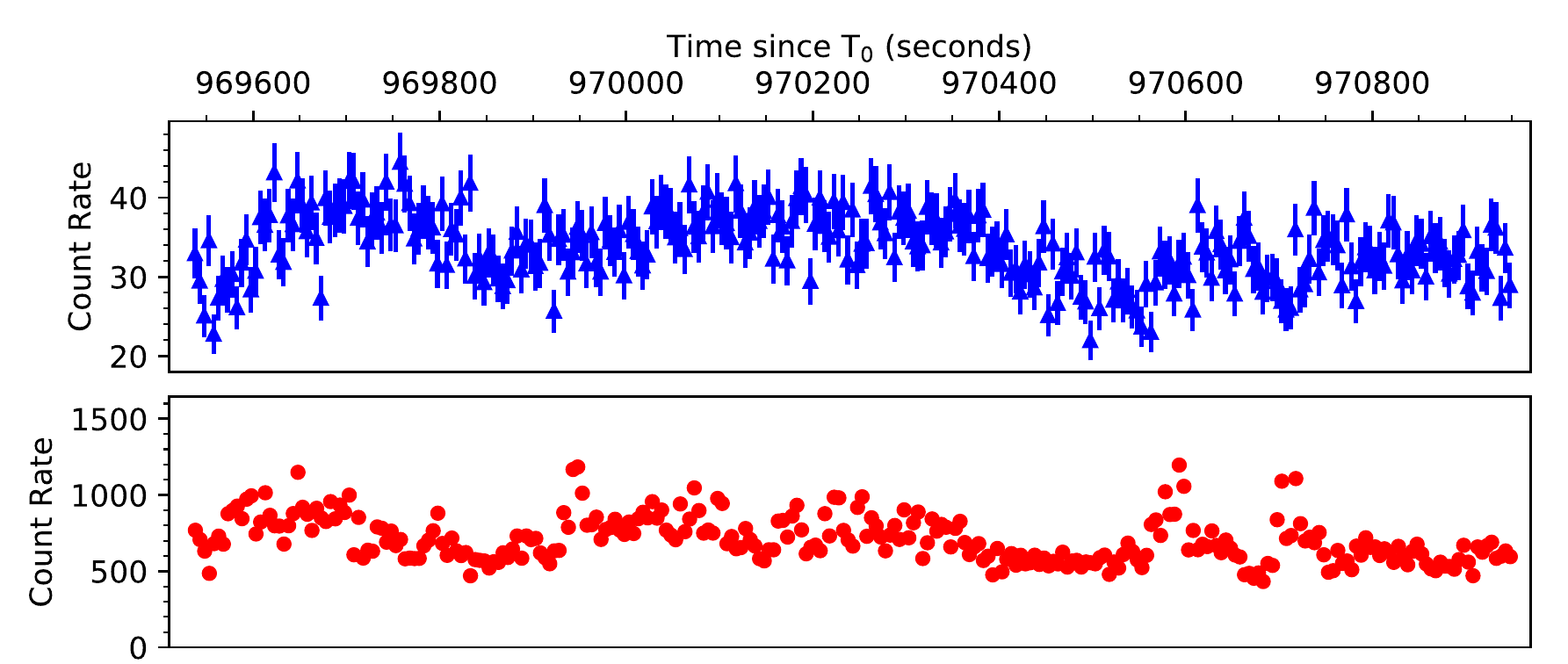}
\includegraphics[angle=0,scale=0.45]{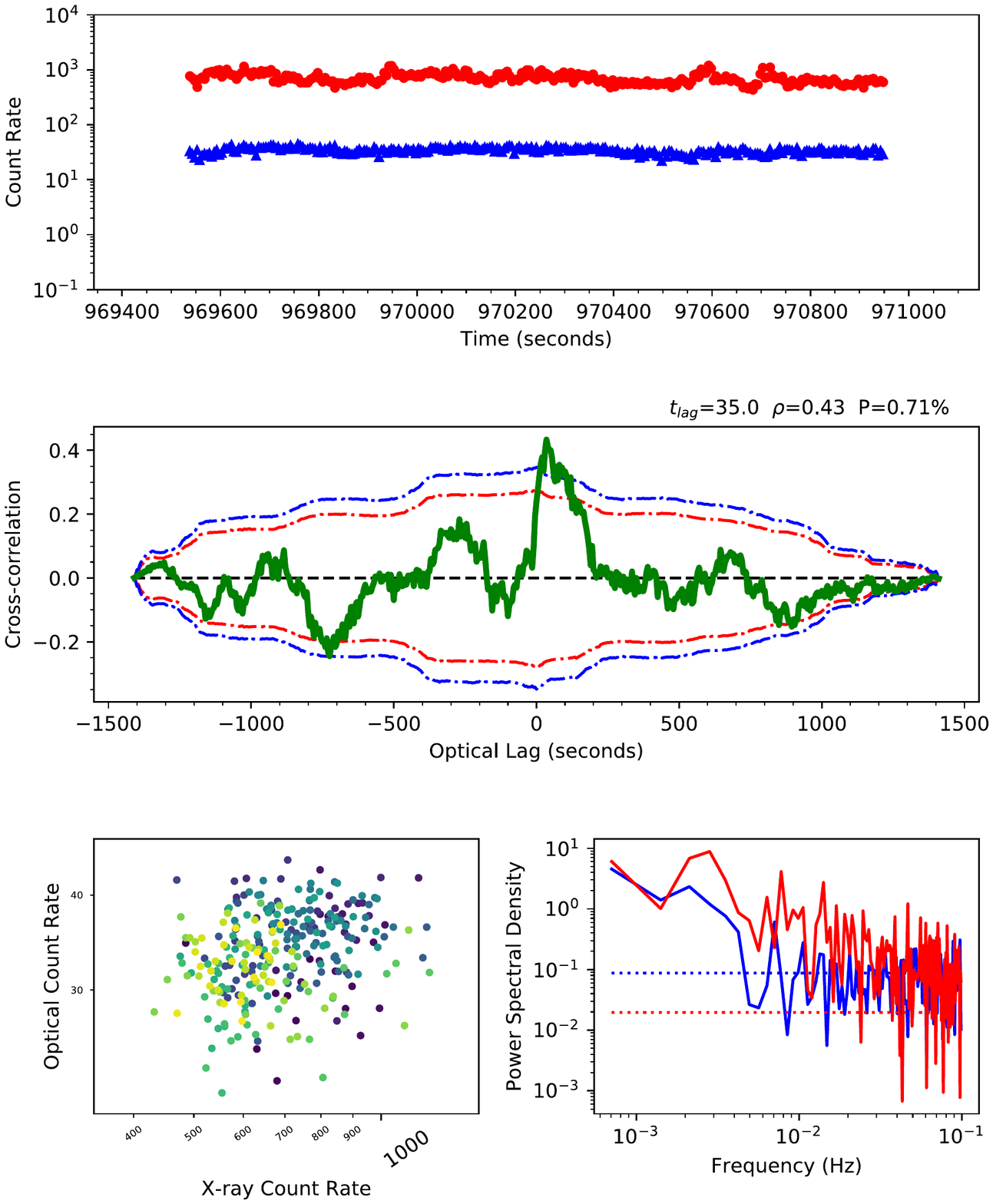}
\centering
\includegraphics[angle=0,scale=0.45]{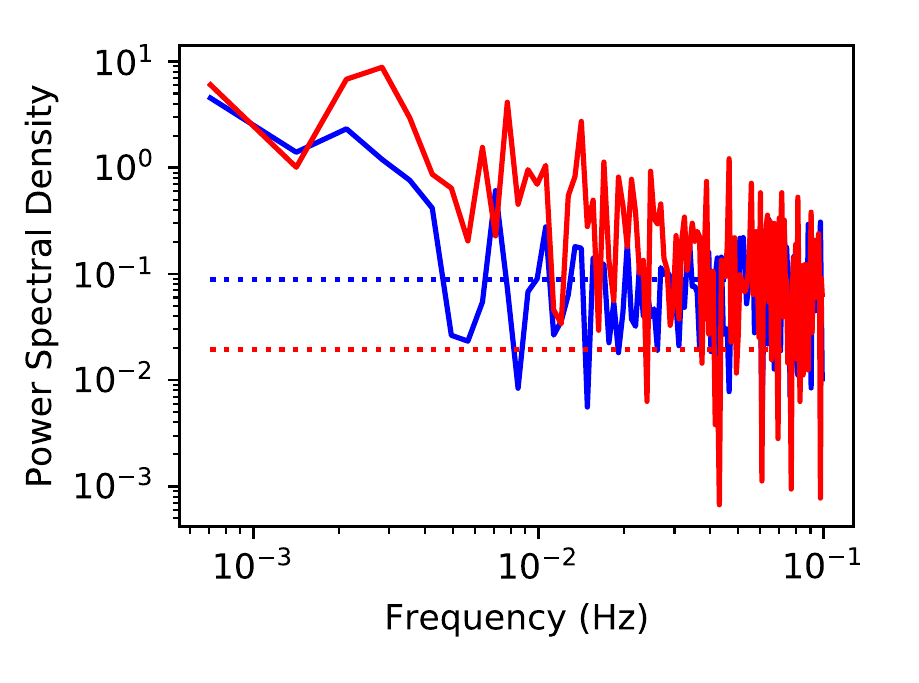}

\caption{Top panel: time sliced observation of the $u$-band (blue triangles) and X-ray (0.5-10\,keV; red circles) observations of V404 Cygni at ${\rm T_0}$+969\,ks with 5\,s binning. Middle panel: the corresponding cross-correlation determined for a range of optical lag times. The blue and red dotted lines indicate the $0.5, 2.5, 97.5$ and $99.5$ per cent percentile boundaries. Bottom Panel: The power spectral density (PSD) for the ${\rm T_0}$+969\,ks segment displaying the PSD for both the X-ray (red line) and $u$-band (blue line) light curves. The dotted lines indicating the expected constant Poissonian white noise contribution. Above the top right corner of the middle panel we give the optical lag time and coefficient at the peak of the cross correlation function and the probability of it occurring by chance considering the multiple cross correlations performed for that segment. The optical and X-ray light curves in the top panel appear to show similar behaviour. This is confirmed by the cross correlation, which indicates the optical and X-ray are correlated with a coefficient of $\rho = 0.43$, but that the optical lags the X-ray emission by 35s. The power spectral density shows that the variability in both bands is due to a combination of red noise, producing the power-law behaviour and Poisson noise, dominating the signal above $>10^{-2}$\,Hz. Similar figures for the full sample are provided in the online material S.2.} 
\label{cross_corr}
\end{figure}

\subsection{High Time Resolution}
\label{shortvar}
In Fig. \ref{eventlcXU}, we show examples of the $u$-band in comparison to the $0.5-10$\,keV X-ray light curves. The full sample can be observed in the on-line material, Fig. S.1. We also show the data again in Fig S.2, but with each panel having the same y-axis and the same time range, allowing an easy comparison of how the behaviour changes from segment to segment. From these figures it is clear that the X-ray base flux changes strongly between segments, while the optical remains fairly constant. The segments cover the period where the outburst is in the brightest phase, i.e., $\rm T_0+\,240$\,ks $-$ 970\,ks ($\rm T_0+\,2.8-11.2$ days). Within each segment the X-rays in general are more variable than the optical/UV: the count rate of the X-ray light curves varies by a factor of up to $\sim$100, whereas the optical count rate changes less drastically by a factor of a few. There are, however, some segments where the optical/UV variability is larger than the X-ray, but the factor is $\lesssim 2$.

On the whole, the $u$-band behaves similarly to the X-ray light curve, but the changes are less prominent and tend to be smoother. Some light curve segments are flat with small variations of a few per cent for both the X-ray and $u$-band light curves, e.g., ${\rm T_0}+579$\,ks and ${\rm T_0}+734$\,ks. Other segments vary more strongly either in one or both bands, with variations on 10\,s and 100\,s time scales, which often appear oscillatory, e.g., segments ${\rm T_0}+$729\,ks, 780\,ks, 792\,ks and 798\,ks. In addition, there are some segments, e.g., ${\rm T_0}+$672\,ks, 820\,ks and 877\,ks, where the X-ray and $u$-band show larger scale variations over the course of a few hundred seconds, but the behaviour is more pronounced in one of the bands. 

We computed the cross correlation for each of the continuous sections of the $u$-band and X-ray light curves. An example is displayed in Fig. \ref{cross_corr} and the entire sample is given in the on-line material Fig. S.2. In these panels, we also show the coefficient values found at the $0.5, 2.5, 97.5$ and $99.5$ per cent level from the simulated coefficient distributions. We find 11 segments to have coefficients which lie outside of either the 0.5 or 99.5 per cent percentile boundaries. Of the 23 segments we tested for a correlation, we would expect by chance that $<1$ of these segments will have a peak coefficient outside of this percentile boundary. However, this statistic should also take in to account that within each segment we are performing multiple cross correlations, one at each lag. Therefore for each segment, we take the chance probability associated with the peak of the cross correlation to be the fraction of simulations that have absolute values that exceed the absolute value of this coefficient at {\it any} lag time. Of the 23 segments, we find 2 segments (segments ${\rm T_0}+240$\,ks and $+969$\,ks given in Fig. S.2) have $<1\%$ chance of the peak cross correlation coefficient being obtained by chance. These segments are a subset of the 11 which have cross correlation coefficients which lie outside of either the 0.5 or 99.5 per cent percentile boundaries. These 2 segments have coefficients that are positive with values of $\rho =0.43$ and 0.64, suggesting the two segments are correlated with slightly different degrees of strength. The optical time lag of these peak coefficient values are 35 and 15s, respectively. In Fig. \ref{cross_corr} and Fig. S.2, in the top right above each optical lag panel we give the optical lag time and coefficient at the peak of the cross correlation function and the probability of it occurring by chance considering the multiple cross correlations performed for that segment. The probability is given only when the value is less than 10 per cent.

\subsubsection{Power Spectra}
For segments of the X-ray and $u$-band light curves of duration $>500$\,s (100 individual data points) we compute the power spectral densities (PSD) from the modulus-square of the discrete Fourier transforms, with the constant of the PSD determined using the usual rms-squared standardisation \citep[e.g.,][]{gan10,utt14}. One example can be observed in the last panel of Fig. \ref{cross_corr}, while the full sample can be observed in the on-line material, Fig. S.2. In the panels displaying the power spectra, we also provide lines indicating the expected constant Poissonian white noise contribution for both the X-ray and $u$-band, which were computed following the procedure in \cite{vau03} and \cite{gan10}. The power spectra appear to be featureless except for a constant and a power-law component. For many of the power spectra, the Poisson noise dominates the signal above $>10^{-2}$\,Hz. Red noise, identified by the power-law slope of $\sim-2$, is the other strong component of the power spectra. However the strength of this component varies and is not apparent in all light curve segments. Steep red noise power spectra were also reported during the outburst of V404 Cygni by \cite{ATel:7677}, \cite{ATel:7688} and \cite{gan16}.

\subsubsection{RMS variability}
To quantify the variability, we computed the excess rms for the high time resolution $u$-band and $0.5-10$\,keV X-ray light curves (see \S~\ref{timing}). The results are shown in the bottom panel of Fig. \ref{v404lc_multipanel}. The excess rms for both the optical/UV and X-ray light curves changes with brightness, increasing as V404 Cygni becomes brighter, with the rms being a factor of 100 larger for the X-ray light curves in comparison to the optical at any given time. In the top panel of Fig. \ref{rms_flux}, we display the square root of the excess rms (rms amplitude) against the mean count rate\footnote{Since the PSDs are a result of a single time series. The measured PSD will therefore not be the true PSD, but will be scattered around it \citep[see e.g][]{vau03}. As a result, the fractional rms values are typically averaged, but due to the highly variable X-ray absorption, the same flux values do not necessarily imply the same spectral properties. Therefore we have chosen not to provide the average fractional rms.}. Both for the $u$-band and X-ray $0.5-10$\,keV data, the excess rms increases with increasing count rate. The dependence on the flux is removed by normalizing the excess rms by the count rate. This results in the fractional variability, which is shown in the bottom panel of Fig. \ref{rms_flux}. 

To quantify the correlation between excess rms and count rate, a function should be fitted to the data, however, we do not expect to obtain a good $\chi^2$ due to the intrinsic scatter of the data. Instead we use a Spearman rank correlation to determine if the logarithms of the excess rms and count rate are correlated. For the X-ray observations we find evidence for a strong correlation with a correlation coefficient of 0.94 and reject the null hypothesis with $p\ll 10^{-5}$. However, for the optical/UV the correlation coefficient indicates a weak correlation, much weaker than for the X-ray. The coefficient is 0.7 and we can not reject the null hypothesis ($p = 0.03$). Furthermore, there are some X-ray data points that imply over 100\% fractional variability. These coincide with segments where the variability spans the entire segment. 

\subsubsection{X-ray and $u$-band count rate comparison}
\label{OXcomparison}
As discussed in \S \ref{shortvar}, the variability in the $u$-band is less pronounced compared with the X-ray. We can explore this behaviour further by displaying all the high time resolution $u$-band and X-ray data together. In the top panel of Fig. \ref{OXcorr}, we show the $u$-band count rate versus the X-ray count rate between ${\rm T_0}$ and ${\rm T_0}+70$ days (0 to $\sim 6000$\~ks). Neither the $u$-band nor the X-ray count rates have been corrected for extinction or absorption and caution should be used at low optical count rate because of the contaminating star (see \S~\ref{UVOTreduction}). At $0.1$ count per second, this nearby star may contribute up to $50\%$ of the total counts, but this fraction decreases as V404 Cygni brightens. The $u$-band and X-ray observations in general increase together, as also shown in Fig. \ref{v404lc_multipanel}, but as V404 Cygni brightens the distribution becomes broader in the X-rays compared to the optical/UV. 

We can look at time-resolved versions of Fig. \ref{OXcorr}, comparing the $u$-band and X-ray count rates of individual light curve segments. These are given in the bottom left sub-panels of the on-line material Fig. S.2. Their behaviour can be divided into two groups: 1) randomly distributed data points, suggesting no strong correlation; these tend to be sections of the light curve where there is no strong variability in either the X-ray or $u$-band; 2) correlated count rates, increasing and decreasing together, although the scatter tends to be large. For two segments the behaviour differs from these two categories: in the ${\rm T_0}$+877\,ks segment the X-ray and $u$-band count rates change independently producing vertical and horizontal lines, while in the ${\rm T_0}$+820\,ks segment similar behaviour is observed, but a hysteresis-like pattern is produced. Large variations in X-ray flux, but no pronounced optical response was also reported by \cite{hyn19}. 

We can examine the relationship between optical/UV and X-rays in luminosity space. The luminosity ratio is shown in the bottom panel of Fig. \ref{OXcorr}, with the colour of each point representing the time of the observation. In this panel, due to the need to create X-ray spectra with enough signal so that spectra could be fitted and a count rate to flux conversion factor could be determined (see \S \ref{XRTreduction}), the data points are coarser in time in comparison to the top panel. In Figure \ref{OXcorr}, the data points are distributed either side of the line of equality, suggesting that there are periods where the $u$-band is more luminous than the X-ray and vice versa. The colour of the points represent the time since the trigger and the colours roughly divide the figure in to horizontal bands. The data at the earliest and latest epochs form roughly horizontal bands at lower optical luminosity. There is weak indication that at low X-ray luminosity ($<$ a few $10^{36}$erg s$^{-1}$) the optical and X-ray may increase at a different rate than at higher X-ray count rate, perhaps resembling the track given in the ${\rm T_0}$+820\,ks or ${\rm T_0}$+877\,ks segments in the on-line material Fig. S.2.

\subsubsection{X-ray and Optical/UV Count Rate Distributions}
\label{OXdistributions}
Another view of the behaviour of the optical and X-ray data can be obtained by examining the distribution of the count rates for each segment of the light curve between $\rm T_0+\,240$\,ks and 970\,ks (${\rm T_0}+\,2.8$ days and ${\rm T_0}+11.2$ days) and for this portion of the light curve as a whole. We compute the logarithmic count rate distribution in the optical/UV and X-ray bands for each segment, binning them dynamically with 20 data points per bin, and express it as a probability density function (PDF), following the methodology of \cite{utt05}. We then checked the consistency of the PDFs with a Gaussian distribution. For 5 separate optical/UV and X-ray segments we discard their PDFs as they contain 3 bins or less. For the 18 remaining segments, we find that 15 and 16 optical and X-ray segments respectively, can be well fitted by a Gaussian distribution, rejecting the null hypothesis for the remaining 3 optical ($\rm T_0+\,672$\,ks, $820$\,ks and $877$\,ks) and 2 X-ray ($\rm T_0+\,797$\,ks and $877$\,ks) segments with $p<10^{-3}$ from the $\chi^2$-test. 

We can also examine the logarithmic count rate distribution during the entire period from $\rm T_0+\,240$\,ks $-$ 970\,ks (${\rm T_0}+\,2.8$ to ${\rm T_0}+11.2$ days), scaling individual X-ray and optical/UV segments by their mean count rate, before computing the PDF. We bin them dynamically with 100 data points per bin, following the methodology of \cite{utt05}. We first explore whether the distributions are consistent with a single Gaussian. We do not find the optical/UV or X-ray PDFs to be consistent with a Gaussian ($\chi_{O}^2/dof=$ 570/18 and $\chi_{X}^2/dof=$ 679/18), but the distributions are singularly peaked. We therefore also attempted to fit a model with 2 Gaussians distributions. This model substantially improved the fit for both the optical/UV and X-ray distributions, resulting in $\chi^2/dof=$ 16/15 and 24/15 for optical/UV and X-ray, respectively.

\begin{figure}
  \vspace{-0.4cm}\includegraphics[angle=0,scale=0.45]{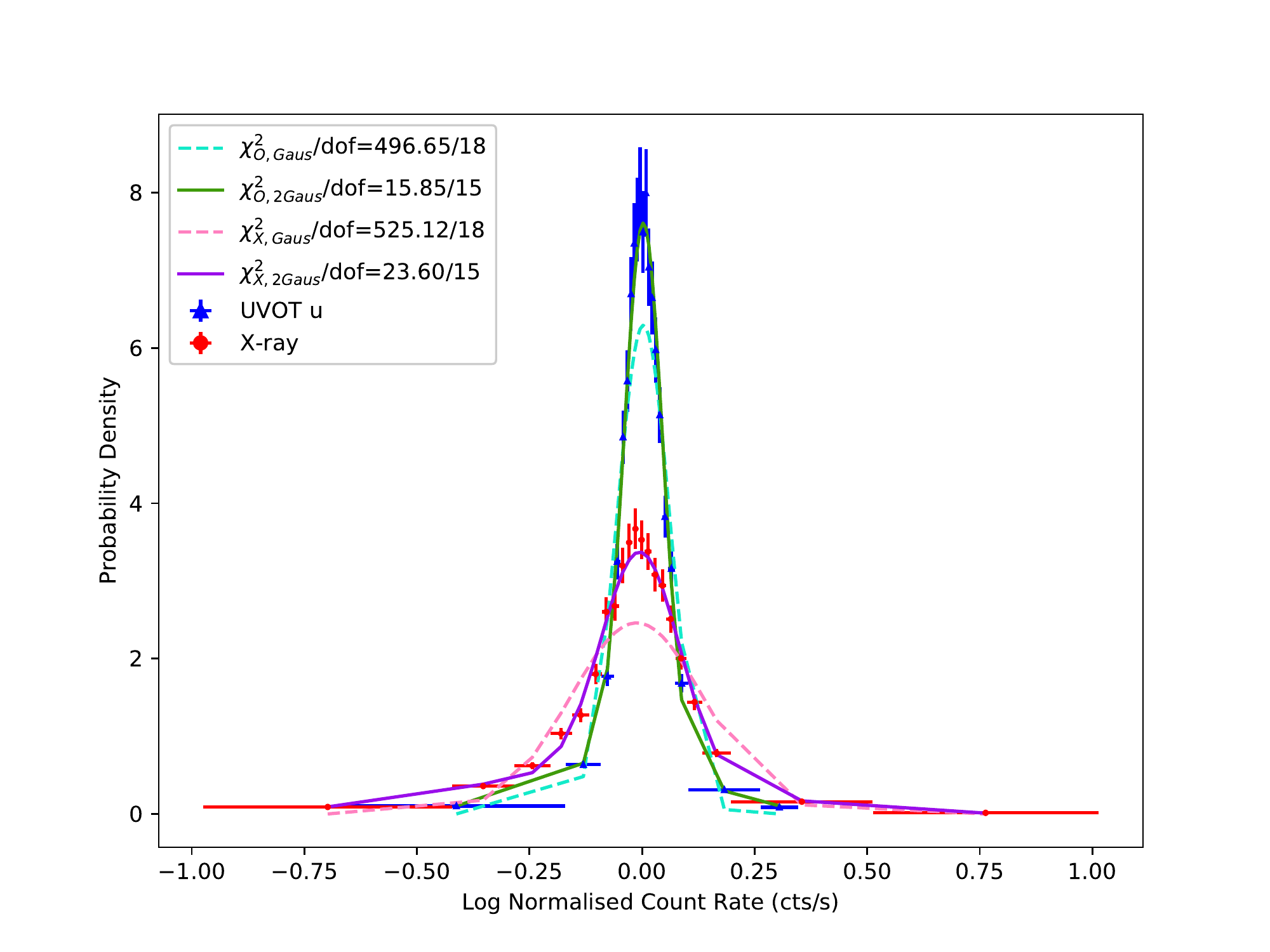}
  \caption{The logarithmic count rate distributions of the outburst between $\rm T_0+\,240$\,ks and 970\,ks (${\rm T_0}+\,2.8$ days and 11.2 days) for the X-ray (red circles) and optical/UV (blue triangles), expressed as probability density functions. The data has been dynamically binned with 100 data points per bin. The dotted lines show the best fit Gaussian distributions ($u$-band: light blue, X-ray: pink). The solid lines show the best fit 2-Gaussian function for each data set (X-ray: purple, $u$-band: green). The count rates from each segment included in the distribution have been normalized by the mean count rate in that segment. The error on each bin is given as the $\sqrt{N}$, where N is the number of count rates in the bin. The $\chi^2$ over the degrees of freedom (dof) for the Gaussian fits are given in the top right corner of each panel.}
\label{flux_dist}
\end{figure}

In Fig. \ref{flux_dist}, we display the lognormal normalized count rate distributions for the $u$-band and X-ray observations, comparing the single Gaussian with the best fit 2 Gaussian models for each data set.
 
\begin{figure}
  \vspace{-0.5cm}\includegraphics[angle=0,scale=0.45]{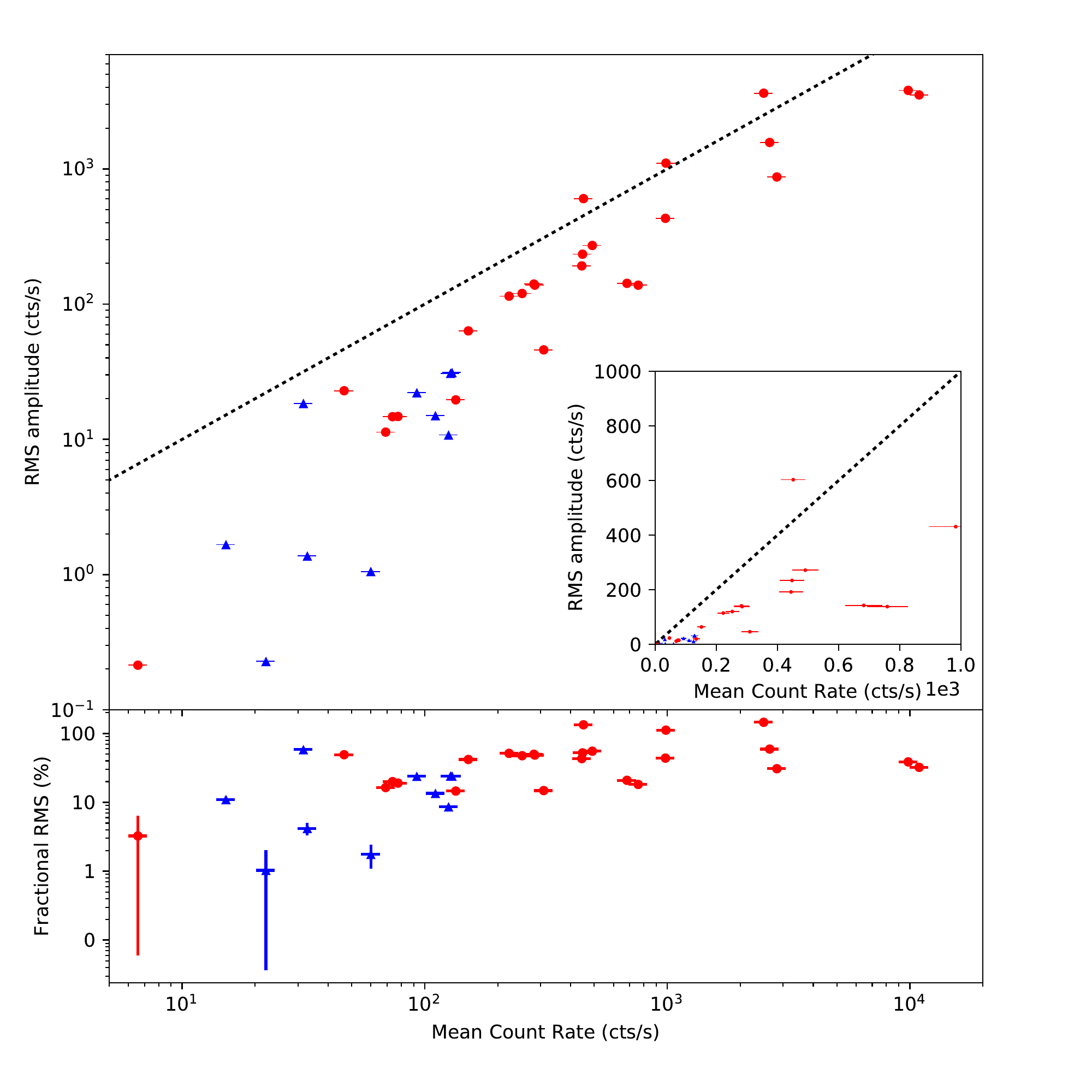}
  \caption{Top Panel: Square root of the excess variance (RMS amplitude) versus the average count rate in log-log space. The dotted line represents the line of equality. The inset panel shows the data up to a count rate of 1000 cts/s in linear space. Bottom Panel: fractional root mean square of the excess variance (fractional RMS) versus the average count rate. Red Circles: $0.5-10$\,keV X-rays, Blue Triangles: $u$-band. All values are determined from 128 consecutive data points of duration 5s. The top panel suggests that for both the $u$-band and X-ray $0.5-10$\,keV data, the amplitude of the variability increases with brightness. This is confirmed in the bottom panel by normalizing the amplitude by brightness and therefore removing this trend.}
\label{rms_flux}
\end{figure} 

\section{Discussion}
\label{discussion}

We have explored and compared the optical/UV and X-ray behaviour of the June 2015 outburst of V404 Cygni. Overall the optical/UV filter light curves and the soft and hard X-ray light curves show similar behaviour. The main outburst consists of a series of large flares on time-scales of a few hours, but strong variations can also be observed at shorter time-scales from a few seconds to hundreds of seconds. The variations are most prominent in the X-rays, while in the optical/UV, they tend to be less prominent and smoother. The X-ray and optical/UV rms-flux ratio indicates that the strength of the variability in both bands is correlated with brightness, but less strongly in the optical. We have shown that strong changes in absorption, as inferred from X-ray spectral fits, do not correspond to changes in optical/UV extinction.

In the following we will first discuss and compare the extinction and absorption, then discuss the rms-flux relation and finally the origin of the variability in the X-ray and optical/UV light curves. We will draw upon all these aspects to provide an overall picture describing the variability observed within V404 Cygni by {\it Swift}.

\begin{figure}
  \begin{center}
      \includegraphics[angle=0,scale=0.4]{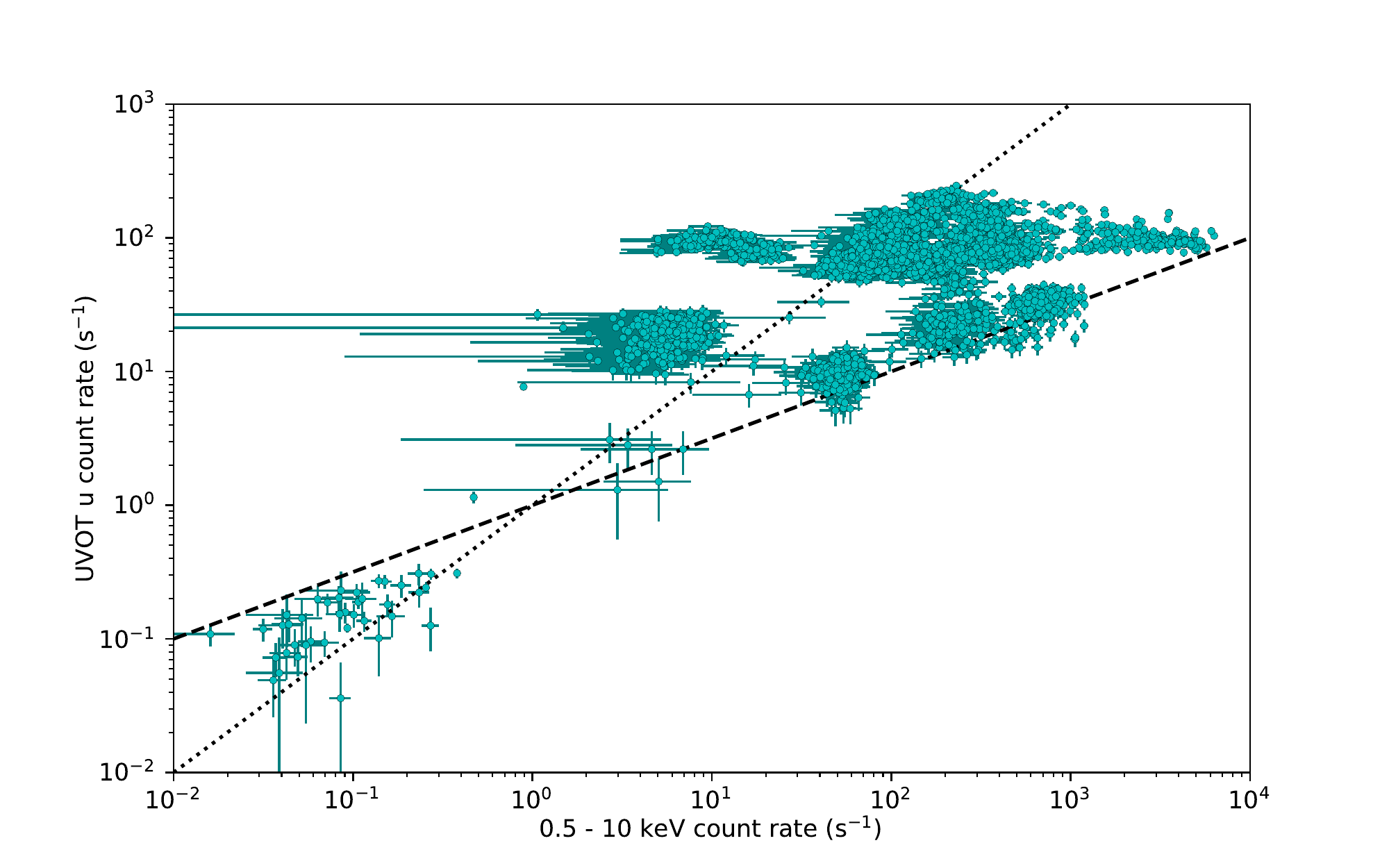}
    \includegraphics[angle=0,scale=0.45]{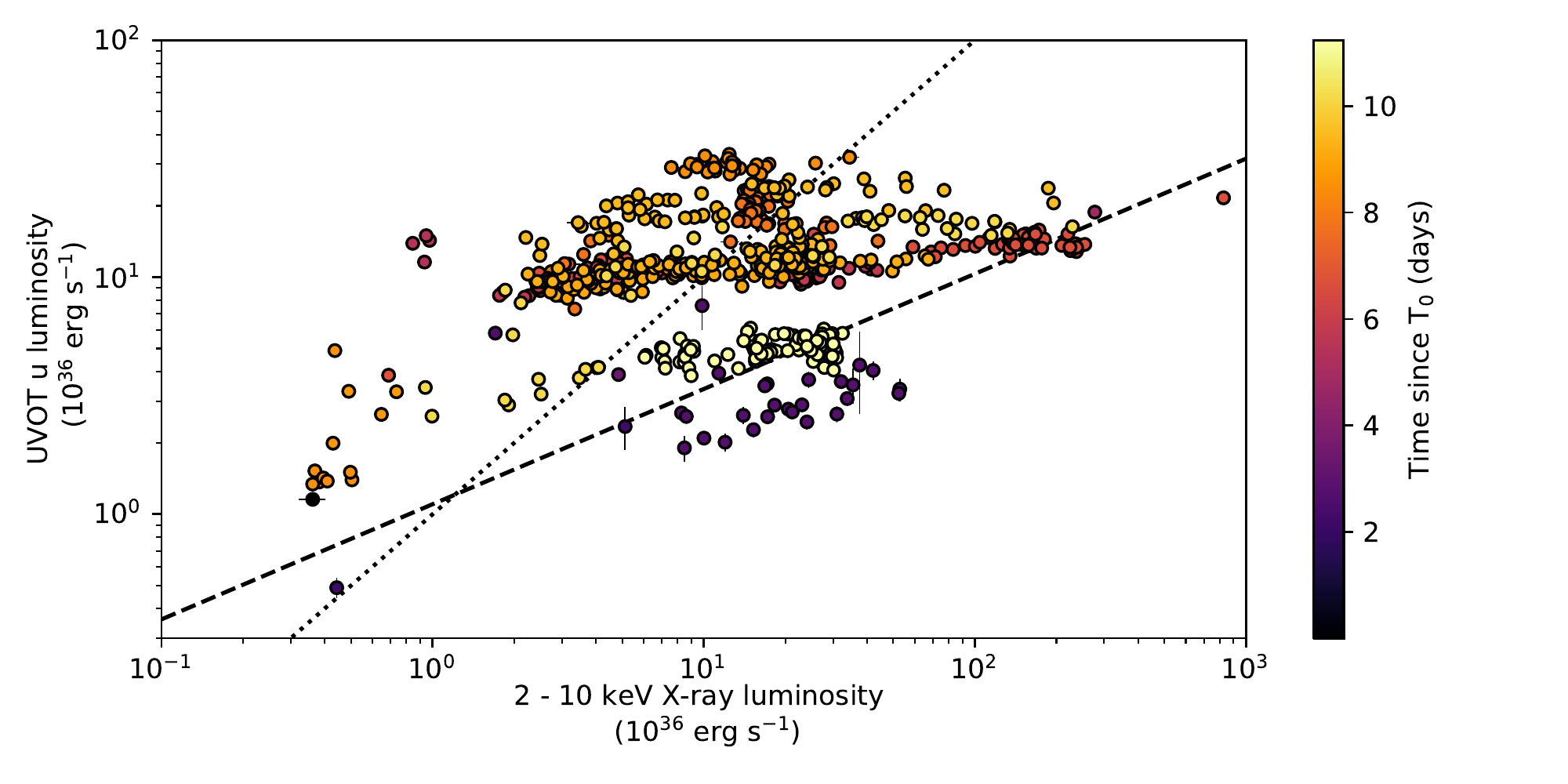}
    \caption{Comparison of the optical and X-ray brightness during the June 2015 outburst. The top panel displays the $0.5-10$\,keV X-ray count rate versus the $u$ band count rate from ${\rm T_0}$ to ${\rm T_0}+70$ days (0 - 6000\,ks). The bottom panel displays the $2-10$\,keV X-ray luminosity together with the UVOT {\it u}-band luminosity. The luminosities have been corrected for Galactic extinction and absorption. The colour represents the time of observation in days since trigger. The luminosity panel displays only X-ray data taken in WT mode since the signal to noise of the PC mode spectra is insufficient to constrain the flux conversion factor. In both panels, the black lines indicate where the $u$-band and X-ray count rate or luminosity are linearly proportional (dashed), and the square root of the X-ray is proportional to the optical (dotted). In the top panel, the $u$-band and X-ray count rates in general increase together, but as V404 Cygni brightens, the distribution becomes broader in the X-rays compared to the optical/UV. This is also observed in luminosity space given in the bottom panel.}
    \label{OXcorr}
  \end{center}
\end{figure}

\subsection{Extinction and absorption}
\label{extincabsorb}
The absorption, as expressed by the column depth $N_H$ observed in the X-ray spectra, changes by over two orders of magnitude. If the X-ray absorbing material contained a significant amount of dust affecting the optical/UV emission, we would expect to see a colour change of $>1.5$ magnitudes. If the absorption and extinction were correlated as found in the Milky Way, the data would folow the dashed black line in Fig. \ref{colour-NH}. However, in Fig \ref{colour-NH} there appears to be no correlation between colour and $N_H$. 

\cite{motta17} suggest that there may be two absorbers: an inhomogeneous high density absorber and a uniform tenuous absorber, which either coexist or the high density rarefies into the lower, more uniform absorber (although rarefaction is less likely due to the fast transitions between the two phases). The separation of the two absorbers cannot be determined through the X-ray observations alone. From our observations, we can rule out the possibility that the optical/UV is affected by the inhomogeneous high density absorber due to the lack of corresponding colour changes and lack of correlated variability during the increase in the X-ray emission associated with sharp drops in the $N_H$, the latter implying the absorber must not be optically thick. Colour evolution was also not observed during a 20 min observation of an optical polarization flare, which ruled out intrinsic extinction as the cause \citep{sha16}. 

There are several possibilities why large extinction is not observed to be associated with the inhomogeneous high density absorber. It may be that the absorber either resides in-between the regions emitting the optical/UV and X-ray emission regions, or it is not large enough to cover the entire optical emitting region.  Alternatively (or in addition), the absorber may be dust-free/have a very low dust-to-gas ratio, or have a flat distribution of dust grain size that results in a flat extinction curve. The dust-to-gas ratio, $E(B-V)/N_H$ must be of order $10^4$ smaller than that typically observed in the Milky Way, in order to not show significant change in colour as the absorption increases by a factor of 1000. To be dust-free the absorber must be within or close to the dust sublimation radius. Taking the quiescent X-ray luminosity for V404 Cygni of $8\times 10^{32}{\rm erg\,s}^{-1}$ \citep{plo17,ber14}, a dust sublimation temperature of 1500K and applying the formalism for the sublimation radius of AGN \citep{nen08a,nen08}, we obtain a sublimation radius of V404 Cygni in quiescence of $\sim10^{12}\,{\rm cm}$, consistent with the size of the accretion disk \citep{zyc99}. Absorption systems entering this radius from the interstellar medium or formed within this radius are likely to have their dust destroyed or are unable to form dust in the first place. The absorber is therefore unlikely to be dusty and thus we do not expect a correlation between absorption and extinction. Additionally, the second, thin absorber, is not ionized so it could correspond to the neutral out-flowing material launched as a disk wind $\sim10^{10}$\,cm from the central black hole \citep{mun16}. Since this region is also within the dust sublimation zone, this wind is unlikely to be dusty and therefore would not strongly affect the optical/UV emission. 

Since the absorbers are not dusty, we may be able to gain some understanding of where the high density absorber lies relative to the regions producing the X-ray and optical/UV emission, assuming the optical/UV emission is produced from the reprocessed X-rays (see the following section for discussion of the origin of the optical/UV emission). We can examine two types of behaviour. The $\rm{T}_0+820$\,ks segment shows strong correlated changes in the optical/UV and X-ray light curves and has time resolved X-ray spectra. In this epoch, an X-ray flare of duration $\sim100$\,s is observed on top of longer term undulating behaviour. A flare is not observed in the optical/UV. The X-ray flare is associated with a rapid decrease in $N_H$, and the covering fraction during this period initially decreases before changing to 100 per cent coverage. In another instance, the two sections ${\rm T}_0+574$\,ks and ${\rm T_0}+579$\,ks, have a constant optical/UV flux level, but the X-ray flux changes by over an order of magnitude. In these instances a decrease in column density appears to be correlated with increase in X-ray flux. Both these types of behaviour suggest that the X-ray source is uncovered and is visible along our line of sight. If the absorber lies in-between the X-ray emitting region and the region where the X-rays are reprocessed to optical/UV, we would expect to see corresponding changes in the optical/UV brightness. While we do find optical and X-ray light curves at times to be correlated, we don't observe the overall optical brightness to change between the ${\rm T}_0+574$\,ks and ${\rm T}_0+579$\,ks epochs. We suggest instead, that the high density absorber is large and lies at a significant distance away from the disk (so that the X-rays are only absorbed after the reprocessing of X-rays to optical/UV; but at this distance, it may already be cool enough for dust to exist - which would lead to extinction), or more likely that the absorber is small and close to the black hole, so that while the central object is uncovered along our line of sight, no additional X-ray flux is transferred to the region producing the optical emission. This is consistent with the schematic view of V404 Cygni given in \cite{motta17}. 

\subsection{RMS-flux relation}
The magnitude of variability and the count rate of both the X-ray and $u$-band light curves scale linearly. Similar behaviour was found in X-ray observations of both GX 339-4 during outburst \citep{mun11} and Cyg X-1 \citep{gle04}, and was initially explored in X-ray binaries by \cite{utt01}. Linear behaviour was also observed in the rms-flux correlations in both the X-ray and optical for a sample of 4 X-ray binaries \citep{gan09}. For GX 339-4, the rms was observed to increase with brightness, however, a continuous hysteresis pattern was observed forming a slanted P shape in the rms-flux (log-log) space \citep{mun11}. \cite{mun11} associated the different phases of the rms-flux diagram with specific spectral phases. The spine of the P shape is associated with the hard state identified on hardness intensity diagrams (HIDs) \citep[e.g., see][for a recent review]{bel16}. The curve of the P shape corresponded to the period where a thermal disk black body component was found in some of the spectral fits of GX 339-4 indicating that the XRB had entered the soft state. For Cyg X-1, the distinction between states in the rms-flux diagram was less prominent. \cite{gle04} showed that the soft and hard states both provided linear rms-flux relations, but the slope was slightly smaller for the soft state compared to the hard state. For V404 Cygni, there is a lot of scatter in the rms-flux relation. It cannot be determined if the observation is a combination of linear relations, but the rms-flux behaviour observed in V404 Cygni is most similar to that found for Cyg X-1 and the XRBs reported by \cite{gan09}. 

Information on the state of the outburst can be determined using the X-ray spectra and rms amplitude of V404 Cygni. In XRBs, the hard state is normally characterized by a power-law-shaped energy spectrum with a photon index of $\sim1.6$ ($2-20$\,keV), and a high level of aperioic variability with an rms amplitude above ∼30 per cent \citep{mun11}. The hard state would be characterised by a thermal disk black body component on the left-hand side of the HID \citep{bel16} and almost no variability is seen \citep{mun11}. For V404 Cygni, none of the X-ray spectra require disk black body components \citep{motta17}. They are best fitted during the main outburst by a power-law with photon index $\lesssim 2$ ($0.6-10$\,keV) and the fractional rms, in the X-ray, on the whole ranges between 10 and 60 per cent. V404 Cygni is therefore consistent with being in the hard state during the main outburst, with the observed rms fluctuations and rms-flux correlation associated with the power-law emission. Two common explanations for X-ray emission in the hard state are that it arises from a ‘corona’ of hot electrons, where seed photons from an optically thin accretion disc are up-Comptonized, or that it originates from a jet \citep{mun11}. The outburst of V404 Cygni therefore never likely reached the disk-dominated soft state \citep[see also][]{motta17}. The near Eddington luminosities could be explained by accretion at Eddington/Super-Eddington rates, which may have sustained a surrounding slim disk, but the flux was partly or completely obscured by the inflated disk and its outflow. In this case, the largest flares produced by the source might not be accretion-driven events, but instead the effects of the unveiling of the extremely bright source hidden within the system \citep{motta17}.

In the lower panel of Fig. \ref{rms_flux}, we have 3 X-ray data points with 100 per cent fractional variability. At these epochs the X-ray light curve is dominated by large undulations in the baseline flux level, changes in flux of order 100. In one case (${\rm T_0}+579$\,ks) the undulation is a large flare\footnote{this X-ray flare, peaking at ${\rm T_0}+578.8$\,ks, is not displayed in online material S.1. as there is no overlapping $u$-band event mode data}, which corresponds to a strong decrease in $N_H$, likely associated with the uncovering of the central object. For the other two points with 100 per cent fractional variability, strong changes in $N_H$ are observed, but do not seem to be the only factor driving the strong changes in X-ray flux. Exploring the rest of the rms values (e.g., those with fractional variability $< 100$ per cent), we find strong changes in $N_H$ are only observed during 2 further data points. This indicates that varying $N_H$ is not solely responsible for the observed rms-flux relation. The optical/UV emission has a lower fractional rms, typically of a few to 20 per cent and only weak evidence for a correlation between rms and flux. This suggests that the mechanism producing the X-ray emission does not directly produce the optical emission, and that, perhaps the X-ray reprocessing is modulating the optical/UV variability.

\cite{utt05} discuss the rms-flux relation as being produced by stationary data. For a light curve to be considered as having stationary variability, the mean and variance should not change with time and thus the flux distribution should be consistent with lognormal. In \S \ref{OXdistributions}, we examined the count rate distributions for each light curve segment, finding that a lognormal distribution is consistent with what we observe for the majority of the individual X-ray and optical light curve segments (see \S \ref{OXdistributions},) between $\rm T_0+\,240$\,ks and 970\,ks ($\rm T_0+\,2.8$ days and 11.2 days). This suggests that the fluctuations for these segments are not additive and are unlikely to be caused by many independent emitting regions or due to shot noise \citep{utt05}. Since most of the light curve segments are consistent with having a lognormal count rate distribution, it is likely that the PSDs used to determine the rms values (1 or 2 per light curve section) are in general measuring stable variability. However, there are 3 and 2 optical/UV and X-ray light curve segments, respectively, where the count rate distributions are inconsistent with a lognormal distribution, suggesting non-stationary behaviour for at least some portion of the light curve. Non-stationary behaviour was also shown by \cite{gan17} for V404 Cygni, in cross correlation of optical and X-ray light curves. We are also unable to state if the variability is stable on much shorter timescales due to the resolution of the data. Furthermore, examining the normalized count rate distribution of the entire time period, we note that the logarithmic X-ray and optical/UV distributions are not consistent with a single Gaussian. This may suggest that the emission process does not result in a logarithmic flux distribution that is Gaussian, or that the emission in the two bands may be produced by more than one process \citep[see also ][]{als19}.

\subsection{X-ray and optical variability}
Within the {\it Swift} high time resolution observations the variability shows three main characteristics: 1) fluctuations on tens of seconds up to a couple of hundred seconds time scales, observed in both bands, but more prominent in the X-ray; 2) larger variations changing the flux level over hundreds of seconds (e.g., ${\rm T_0}+877$\,ks, see supplementary material Fig. S1), observed in both bands, although more strongly in one (typically the X-ray) compared to the other; 3) the base-line count rate of both bands change between orbits/segments of data, e.g., the ratio between the optical and X-ray count rate varies from epoch to epoch, typically the X-ray varies more strongly than the optical/UV. 

The range in behaviour between the optical/UV and X-rays suggests that the variability may not be due to a single mechanism. One possible explanation for the production of the optical/UV variability is reprocessing of the X-ray emission by the disk \citep{gan16, kim16, rod15}.  The X-ray emission originates from the inner regions near to the black hole. Modelling of spectral energy distributions (SEDs) performed by \cite{kim16} suggests reprocessing is present. The high degree of correlation between the $u$-band and X-ray light curves observed in two of our light curve segments supports reprocessing as the dominant source of emission at least for this portion of the outburst. The light curve segment at ${\rm T_0}+$820\,ks shows large scale variability in the X-ray and similar emission in the optical, but that is broader in time compared with the X-ray. This is a classic expectation for reprocessing, where by the lag and blurring of the optical light curve depends on the reprocessing area \citep[e.g.][]{obr02,hyn06}. Therefore, when the optical/UV variability is broader than the X-rays, we can assume that the optical/UV emitting region is larger than the X-ray. \cite{motta17}, using {\it Swift}/XRT and INTEGRAL observations, constrain the X-ray emission to within 100 gravitational radii, with softer X-rays being produced within 10 gravitational radii. \cite{wal17} with NuSTAR observations also suggest that the X-ray emission, at least during the flares, is produced close to the black hole within ∼10 gravitational radii. With a black hole mass of 9 $M_\odot$, it has been estimated that the the separation between the black hole and the companion is $\sim2.2\times10^{12}$ cm or $\sim75$ light seconds \citep{rod15, gan17}. The short time lag observed for the segments of our optical and X-ray light curves, similar to that reported by \cite{ATel:7727}, \cite{kim16} and \cite{alf18}, is consistent with this picture. However, our typical measured optical lag of $<35$ sec, may suggest that the $u$-band emission from reprocessing does not originate at the very outskirts of the disk, but at a radius of $\lesssim1\times10^{12}$ cm. Standard disk reprocessing is expected to produce optical variations that follow an $L_X^{1/2}$ trend \citep{van94}. This relationship is shown in Fig. \ref{OXcorr}, but it is not consistent with the overall behaviour observed in either count rate or luminosity space. In Fig. \ref{OXcorr} horizontal track behaviour is observed with the X-ray luminosity appearing to increase without significant change in the optical/UV, this is also observed in the equivalent count rate figure for the ${\rm T_0}+$877 segment shown in Fig. S.2. This horizontal behaviour could suggest that either the X-ray and optical/UV emission are produced independent or that the optical/UV emitting region is unable to respond to the increase in the X-rays \citep[see also][]{hyn19}. However, the optical/UV count rate versus X-ray count rate figure of the ${\rm T_0}+$877 segment also displays a vertical line, which shows the optical/UV count rate changing while the X-ray count rate remaining constant. This could suggest a scenario where by the production of the optical/UV emission is delayed in comparison to the X-ray perhaps consistent with reprocessing. However, we can not draw strong conclusions from these figures since the X-ray count rates and luminosities are not corrected for intrinsic absorption, which may be responsible in part for the observed horizontal track behaviour in Fig. \ref{OXcorr}.

There is considerable evidence suggesting that reprocessing is not solely responsible for producing the optical/UV and X-ray variability. Lack of X-ray variations coincident with optical variations suggest an alternative mechanism to reprocessing for producing at least some of the optical variations \citep{dal17}. In addition, long lags discovered between the optical and INTEGRAL data by \cite{rod15} from 0\,s up to 20-30\,min \cite[see also][]{alf18} also cannot readily be explained by reprocessing given the size of the binary. Within the {\it Swift} XRT/UVOT observations, we also find that reprocessing may not be responsible for all the observed temporal fluctuations. For instance, in the ${\rm T_0}+$671 $-$ ${\rm T_0}$+672\,ks segment the dip in the optical/UV count rate is stronger than that observed in the X-ray. 

Variable absorption has also been proposed as the cause of at least some of the X-ray variations \citep{kin15, motta17}. \cite{motta17} identified strong spectral variations in the time resolved $0.5-10$\,keV X-ray spectra of V404 Cygni, which they identify as due to variations in the layer of material covering the source. They find that changes to the column density cause moderate changes to the flux on timescales of minutes, but that larger flux variations are due to sudden drops in the local column density and simultaneous disappearance of partial covering. These changes in column density are proposed as the covering/uncovering of a bright central source causing the large scale flux variations. Indeed, strong change in $N_H$ is observed within the ${\rm T_0}+$820\,ks and ${\rm T_0}+$877\,ks segments and between the ${\rm T_0}+$574\,ks and ${\rm T_0}+$579\,ks segments. We do not observe flux variations in the optical/UV that are coincident with changes in the X-ray flux due to varying absorption. This is to be expected as the optical emission is unlikely to be affected by the local absorbers since the dust is likely sublimated (see \S\ref{extincabsorb}). While changing absorption is a cause of X-ray variability in V404 Cygni, it is unlikely to be the sole cause since we observe correlated variable behaviour between the optical and X-ray light curves, even when rapid, short lived changes in flux due to changing $N_H$ are observed in the X-ray, e.g., ${\rm T_0}+820$\,ks and ${\rm T_0}+877$\,ks segments. 

An alternative mechanism for producing the observed X-ray and optical/UV emission and associated variability is a non-thermal jet. Evidence for jet emission in V404 Cygni has been shown in the hard X-rays through to radio wavelength observations \citep[see also][for a list of references for jet emission in the 2015 outburst of V404 Cygni]{mai17}. The earliest evidence for a jet was observed by INTEGRAL between ${\rm T_0}$+2 to ${\rm T_0}$+5 days ($\sim \rm T_0+\,170$\,ks - 430\,ks), with hard emission due to Compton-upscattering of seed photons likely originating in a synchrotron driven jet \citep{nat15,roq15}. At optical/UV wavelengths, \cite{gan16, gan17} propose that the jet was responsible for producing the sub-second optical/UV variability observed by ULTRACAM during the brightest phase of the outburst, while \cite{mai17}, propose that the slow variability is also a result of synchrotron emission from the jet because they see no optical spectral changes over a factor of 20 change in optical flux. They propose that the sub-second variability and slow variability may originate from two different jet components: an optically thick component that produces the slower variations and an optically thin component that exhibits red and rapid variability. \cite{dal17} also find evidence for optical variability being produced by the jet; they focus on a sharp drop in flux in X-ray - IR bands at MJD 57198.23, which lasts $\sim 100$\,s. They exclude a thermal origin for the emission during this decline, ruling out X-ray reprocessing as the source of the optical/IR emission. They instead conclude synchrotron cooling process is the dominant source of the optical/IR emission. \cite{gan17} also find, coincident with an increase in radio brightness, that the optical lags the X-ray by $\sim 0.1$\,s. They conclude that the X-ray and optical emission is attributed to synchrotron emission from the jet, with the optical being produced at a distance $\simeq 10^3$ gravitational radii from the X-ray emission.

It is likely that at least some of the optical/UV variability observed by the UVOT results from the jet, potentially contributing to the optical emission from ${\rm T_0}$+2 days ($\sim \rm T_0+\,170$\,ks) onwards. The dip observed between the ${\rm T_0}+671$\,ks and ${\rm T_0}+672$\,ks segments may be consistent with a picture of varying accretion causing variability within the jet. Examining the colour terms in Fig. \ref{v404lc_multipanel}, we can derive a spectral slope from the $uvw1-u$ colour term, which remains approximately the same during the main outburst, resulting in a spectral index $\beta \sim 0.3$; this value is approximate and is dependent on uncertainties in the reddening correction and flux conversion factor. However, this spectral index is consistent with the spectral index between $V$ and $Ic$ filters observed by \cite{mai17} and the $u'$ and $g'$ filters observed by \cite{gan16}, although the high time resolution spectral energy distributions (SEDs) of \cite{gan16} show that the spectral slope changes significantly, typically between 0-0.5 during their observations on 4 different nights. A spectral index of $1/3$ is consistent with what is expected for a steady-state disk with radiation due to viscous dissipation, but \cite{mai17} favour an optically thick jet, due to the persistence of the spectrum through the large variability and because typical X-ray transients show steeper optical spectra \citep{hyn05}. 
 
Overall, the discrepancies from standard reprocessing and the evidence of a jet, suggest that the variability we observed is built up from a complex amalgamation of processes and conditions that dominate at different times during the outburst. The lack of strong cross-correlation coefficients and a dominating lag time supports our conclusion requiring the outburst to be a result of multiple processes that may be contributing at different levels during different segments of light curve. This will likely reduce or at least complicate any potential cross-correlation signal. 

\section{Conclusions}
\label{conclusions}

We have presented observations of the 2015 outburst of V404 Cygni, performed by {\it Swift}. We examined observations taken simultaneously by the UVOT and XRT in the optical/UV and X-ray bands, respectively, comparing the properties of the source in these two bands.

We examined the relationship between extinction and (neutral) absorption measured from the optical/UV and X-ray data, respectively, finding no correlation between the two. Due to the lack of corresponding colour changes in the optical/UV band we excluded the possibility that (in contrast to the X- ray emission) the optical is affected by an inhomogeneous high density absorber. We conclude that the high density absorber is likely small, contains a negligible amount of dust and resides close to the black hole. 

We also analysed the temporal behaviour of the optical/UV and the X-ray emission. We found that strong variations can be observed on tens of seconds to hourly time-scales. The variations are most prominent in the X-rays, as opposed to those observed in the optical/UV, which are less prominent and smoother. In both bands the amplitude of the variations are correlated with flux, but to a lesser extent in the optical/UV. We found that within a few outburst phases the optical/UV and X-rays are correlated, with an optical lag typically $15-35$\,s. We conclude that the variability in the optical/UV and X-ray bands is due to a complex combination of processes. Variability visible in both the X-ray and optical/UV bands may be accretion-driven, with either the X-rays produced in the inner accretion disc, which is then reprocessed to the optical/UV, or with both the X-rays and optical/UV produced within the jet. In addition, some of the variability that is observed in the X-ray band only, may be due to the presence of a local, inhomogeneous and dust-free absorber.

\section{Acknowledgements}
We thank the referee for their constructive comments and suggestions. This research has made use of data obtained from the High Energy Astrophysics Science Archive Research Center (HEASARC) and the Leicester Database and Archive Service (LEDAS), provided by NASA's Goddard Space Flight Center and the Department of Physics and Astronomy, Leicester University, UK, respectively. SRO gratefully acknowledges the support of the Leverhulme Trust Early Career Fellowship. AJCT and SRO acknowledge support from the Spanish Ministry, Project Number AYA2012-39727-C03-01. SEM acknowledges the Violette and Samuel Glasstone Research Fellowship program and the UK Science and Technology Facilities Council (STFC) for financial support. PG thanks STFC and UKIERI for support. This work was also supported by the UK Space Agency. We also thank J. Lyman for useful discussion.


\bibliographystyle{mn2e}   
\bibliography{V404cyg_UVOT} 

\IfFileExists{\jobname.bbl}{}
 {\typeout{}
  \typeout{******************************************}
  \typeout{** Please run "bibtex \jobname" to optain}
  \typeout{** the bibliography and then re-run LaTeX}
  \typeout{** twice to fix the references!}
  \typeout{******************************************}
  \typeout{}
 }
 
\newpage
\appendix
\label{appendix}

\label{Dec_app}

\renewcommand\thefigure{A.\arabic{figure}}    

\begin{figure*}
  \begin{center}
    \includegraphics[angle=-90,scale=0.45]{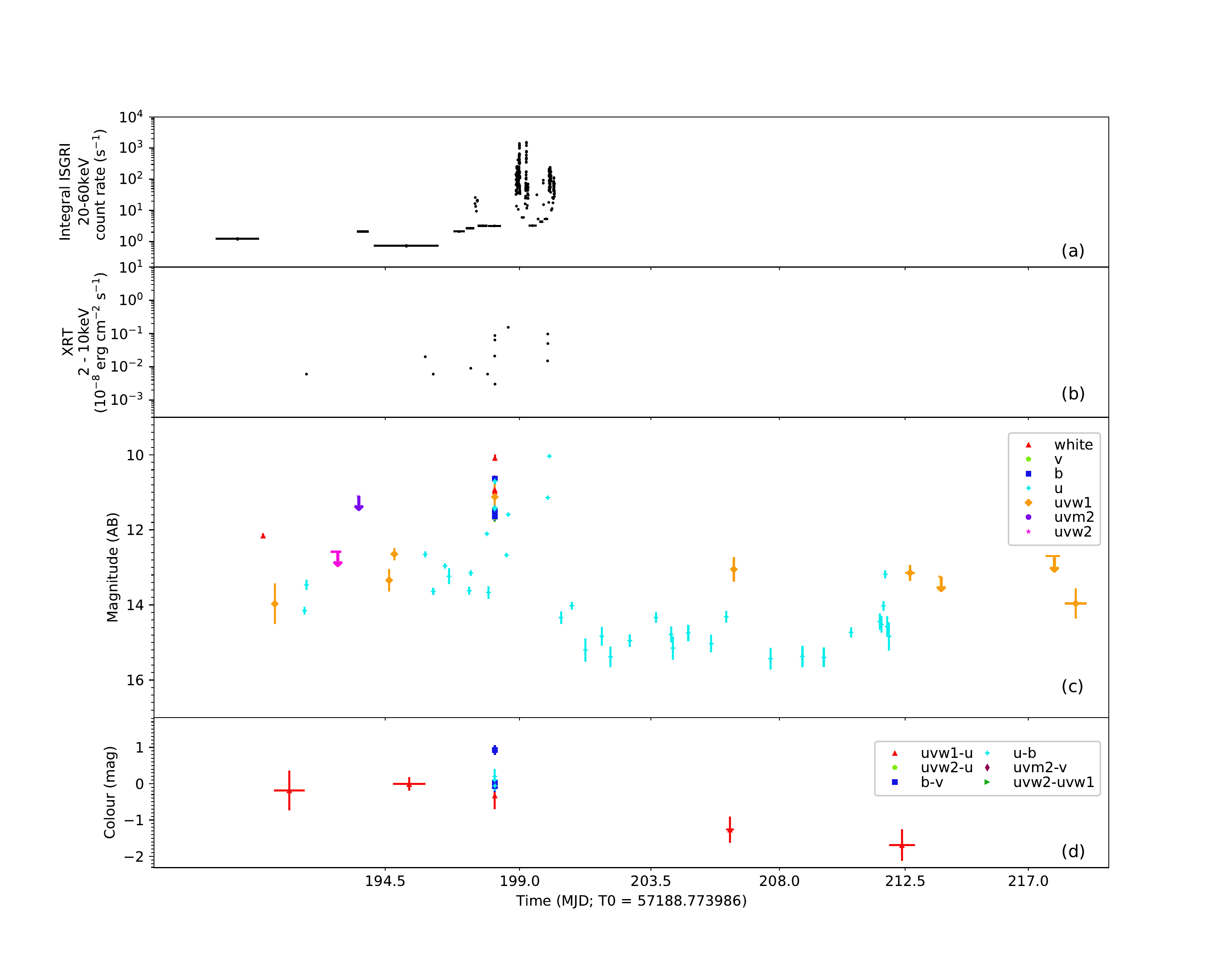}
    \caption{Multi-wavelength observations of V404 Cygni performed during the December 2015 outburst. $\rm T_0$ is taken as the time at which {\it Swift}/BAT first detected V404 Cygni on the 15th June 2015 and has been corrected to the barycenter. Panel a) shows the INTEGRAL 20-60\,keV ISGRI light curve for the December 2015 outburst. Panel b) shows the X-ray $2-10$\,keV flux light curve for the December 2015 outburst. Panel c) displays the 7 optical/UV light curves observed by UVOT. Panel d) displays the colour light curves for several UVOT filters, determined only for points with S/N $>2$. Magnitudes are given in AB and have been corrected for the measured extinction $A_v=4.0$ \citep{cas93,hyn09}. Colours and symbols for each of the UVOT filters and colour light curves are provide in the legends. Downward arrows are $3\sigma$ upper limits computed when the signal to noise is $<2$.}
    \label{Dec2015}
  \end{center}
\end{figure*}

\section{December 2015 Outburst}
At ${\rm T_0}\sim190$ days, {\it Swift}/BAT detected a secondary outburst \citep{GCN:18716}. At this time, {\it Swift} UVOT and XRT resumed observing and continued to monitor V404 Cygni for an additional 30 days, although the sampling was less frequent compared with the June 2015 outburst. For these observations the optical/UV and X-ray data were reduced following the method outlined in \S \ref{reduction}.

The light curves from the UVOT, XRT and INTEGRAL can be observed in Fig. \ref{Dec2015}. At this time, the optical/UV flux level was already above the quiescent level observed prior to 2015 and at the end of the June 2015 outburst. The brightest flux values occurred between $\rm T_0$+199 and ${\rm T_0}+200$ days. After peak brightness, V404 Cygni rapidly decayed, within a day, down towards the pre-2015 level, which it fluctuated around until the end of observations. The X-ray light curves observed by XRT and INTEGRAL behave in a similar way. This mimics the behaviour observed in the June 2015 outburst and the duration of the outburst is similar. However, the outburst is weaker at all frequencies. In the $u$-band the peak magnitude is $\sim 3$ orders of magnitude fainter than the peak magnitude of the June 2015 outburst. This is in agreement with the behaviour observed for the June and December outbursts in i, R, V and B bands reported by \cite{kim17}. Examination of the colour light curves, for which the best sampled for the December outburst is $u-uvw1$, suggests the spectrum becomes bluer from the start of the outburst through to quiescence, although we note that, except for the flare at ${\rm T_0} \sim 198$\,days, the $uvw1$ band is relatively flat in comparison with the $u$ band and the colour change is driven by a change in the $u$-band flux.

\end{document}